\documentclass[pra,amsmath,amssymb,twocolumn]{revtex4}

\usepackage{amsmath}
\usepackage{amssymb}
\usepackage{subfigure}
\usepackage{graphicx}
\usepackage{ulem}
\usepackage{multirow}
\usepackage{natbib}

\newcommand{\be}{\begin{equation}}
\newcommand{\ee}{\end{equation}}

\newcommand{\lf}{\left}
\newcommand{\rg}{\right}
\newcommand{\ra}{\rangle}
\newcommand{\la}{\langle}
\renewcommand{\a}{\alpha}
\newcommand{\bea}{\begin{eqnarray}}
\newcommand{\eea}{\end{eqnarray}}
\newcommand{\cc}{{\cal C}_2}
\renewcommand{\o}{\omega}

\newcommand{\nn}{\nonumber}

\renewcommand{\k}{\kappa}

\newcommand{\D}{\Delta}

\newcommand{\hf}{\frac{1}{2}}

\usepackage{xcolor}

\begin{document}
\title{Non-abelian Thouless pumping in a photonic lattice}
	%\title{Flat bands fractional photon dynamics.\\{Non-abelian holonomies in a  photonic crystal.}\\
		%\title{Non-abelian photon dynamics in a flat band lattice}
		%{\textcolor{red}{Non-abelian holonomies in a generalized Lieb lattice.}}}
	
	\author{Valentina Brosco}%
	\email{valentina.brosco@roma1.infn.it}
	\affiliation{Institute for Complex Systems, National Research Council (ISC-CNR), Via dei Taurini 19, 00185 Rome, Italy}%
		\author{Laura Pilozzi}
	\affiliation{Institute for Complex Systems, National Research Council (ISC-CNR), Via dei Taurini 19, 00185 Rome, Italy}%
        \author{Rosario Fazio}
	\affiliation{The Abdus Salam International Centre for Theoretical Physics, Strada Costiera 11, I-34151 Trieste, Italy 
Dipartimento di Fisica, Universit\`a di Napoli “Federico II”, Monte S. Angelo, I-80126 Napoli, Italy†}%
	\author{Claudio Conti}
	\affiliation{Institute for Complex Systems, National Research Council (ISC-CNR), Via dei Taurini 19, 00185 Rome, Italy}%
	\affiliation{Department of Physics, University Sapienza, Piazzale Aldo Moro 5, 00185 Rome, Italy}%

	\date{\today}
\begin{abstract}
 Non-abelian gauge fields emerge naturally in the description of adiabatically evolving quantum systems having degenerate levels.  
 Here we show that they also play a role in Thouless pumping in the presence of degenerate bands.
 To this end we consider a photonic Lieb lattice having two degenerate non-dispersive modes and show that, when the lattice parameters are slowly modulated, the propagation of the photons bear the fingerprints of the underlying non-abelian gauge structure. The non-dispersive character of the bands enables a high degree of control  on photon propagation. Our work paves the way to the generation and detection of non-abelian gauge fields in photonic and optical lattices.
\end{abstract}


\maketitle

To describe the dynamics of a quantum system whose  Hamiltonian depends on some external classical parameters in the energy eigenstate basis,
%If  $H({\bf \l})$ is a multiparameter family of Hamiltonians
% and $\phi_n({\bf R})$ is an isolated energy eigenstate %whose structure depends on the external parameters, 
one is naturally led to introduce a gauge field \cite{bott1965} that keeps track of how the eigenstates transform in parameters space. 
As first noted by Simon \cite{simon1983}, the Berry's phase \cite{berry1984} is the holonomy associated with this gauge structure when the evolution of the system takes place in an isolated non-degenerate eigenstate.
If the evolution involves a degenerate energy eigenspace the underlying gauge theory becomes non-abelian  \cite{wilczek1984}.\\
%
%
%
%
%The striking relevance and universality of 
%Berry's connection and Berry's curvature \cite{wilczek1989} emerge in many different frameworks ranging from modern polarization theory \cite{resta2007} to topological phases \cite{hasan2010}. %  and  synthetic dimensions \cite{ozawa2019a,kremer2020}.
\noindent Among the most fascinating and studied manifestations of Berry's curvature \cite{wilczek1989,resta2007}  are those that involve transport with the closely related phenomena  of  quantum  Hall effect \cite{thouless1982} and Thouless pumping \cite{thouless1983}. 
%The Berry's phase is the holonomy of this connection 
%
Thouless pumping refers to the transport of charge in the absence of any external bias by cycling adiabatically in time a one-dimensional lattice potential confining the charge.
It is considered as a natural probe of Berry's geometric phase \cite{aunola2003,mottonen2006,xiao2010,zilberberg2018,PhysRevB.93.245113,Aidelsburger2014} and of  the Chern numbers \cite{thouless1983,kraus2012,ke2016,wimmer2017,fedorova2019,wang2019,lohse2015,nakajima2016,kremer2020}.
% Pioneering experiments implementing Thouless pumping in optical lattices loaded with bosonic \cite{lohse2015} and fermionic \cite{nakajima2016} atoms further raised the interest in this phenomenon. 
% %the topological properties of the pump Hamiltonian 
%Recentl pioneering experiments implementing Thouless pumping in optical lattices loaded with bosonic and fermionic atoms further raised the interest in this phenomenon. 
 %the topological properties of the pump Hamiltonian 
%%% 
%%

Here we show that, in the presence of degenerate bands,  Thouless pumping can be exploited to probe the underlying non-abelian gauge structure. 
The non-abelian analogue of the Berry's phase,  %then a matrix-valued unitary transformation in the degenerate subspace, {\sl i.e.}
the holonomy of Wilczek and Zee \cite{wilczek1984}, keeps track of how the degenerate subspace deforms along a 
path in parameters space and it depends on the geometric and topological structure of the Hilbert space. 
We establish a relation between the Wilczek and Zee holonomy and Thouless pumping.

To exemplify our findings we consider a simple one-dimensional lattice model that we name for simplicity non-abelian Lieb chain (naL)  since it is reminiscent of the two-dimensional Lieb lattice model \cite{lieb1989}, recently at the focus of pioneering experiments on light confinement in photonic waveguide lattices \cite{vicencio2015,mukherjee2015}. We show that the slow cyclic modulation of the model parameters yields a current that can be directly connected to the holonomy generated during the cycle.
The non-abelian Lieb chain can be experimentally realized in  a photonic waveguide lattice and the non-abelian holonomy can be read-out from the photon beam displacement along its propagation. Our results thus point at a straightforward approach to generate and detect  Wilczek-Zee non-abelian holonomies. 
In this respect, we remark that, in spite of a few theoretical proposals \cite{duan2001,taddia2017,faoro2003,brosco2008,pachos2000,iadecola2016,chen2019},
only two experimental verifications of Wilczek-Zee non-abelian holonomies exist so far, namely, the pioneering nuclear quadrupole resonance  experiment of Ref.[\onlinecite{zwanziger1990}] and the circuit QED \cite{martinis2020} implementation of Abdumalikov {\sl et al.}\cite{abdumalikov2013}. 

Photonic waveguide lattices are the ideal playground to realize topological lattice models \cite{ozawa2019}.
In these systems, consisting of arrays of evanescently coupled single-mode waveguides \cite{christodoulides2003}, light propagation can be modeled
by means of a Schr\"odinger equation where the role of time is played by the coordinate along the waveguide, $z$, and the Hamiltonian 
describes the hopping of bosonic particles across the array. Such lattice-based description has long been applied to describe light propagation in photonic waveguides arrays and it can be derived using coupled-mode theory, as discussed in various textbooks \cite{snyder} and  briefly outlined in Section \ref{sect-photonics}.
%where
To implement Thouless pumping, the diagonal elements of the lattice Hamiltonian, can be modulated along $z$ by tuning the waveguide diffraction index and the lateral confinement length, while the hopping matrix elements can be controlled by changing  the overlap of the evanescent tails of neighboring waveguides, {\sl e.g.} modifying the lattice spacings. %$n$ and $n^\prime$.
 Recently, modulated photonic waveguides  were employed to implement adiabatic population transfer \cite{PhysRevResearch.1.033117} in a tripod system corresponding to a single unit cell of the naL lattice proposed below.
 
The paper is organized as follows. After introducing  the naL chain model  in Section  \ref{sect-model}, we present our main results concerning the relation between Wilczek-Zee holonomy  and Thouless pumping  in Section \ref{sect-main}, eventually,  in Sections \ref{sect-cycles} and \ref{sect-photonics} we illustrate the results starting from specific examples of pumping cycles. In section \ref{sect-photonics} we outline the derivation of the lattice model equations from coupled mode theory, we  discuss some experimental requirements to discern genuine geometric effects from non-adiabatic transitions.

 \section{Non-abelian Lieb chain}
 \label{sect-model}
\begin{figure}[b]
	\begin{center}
		\includegraphics[width=1\columnwidth]{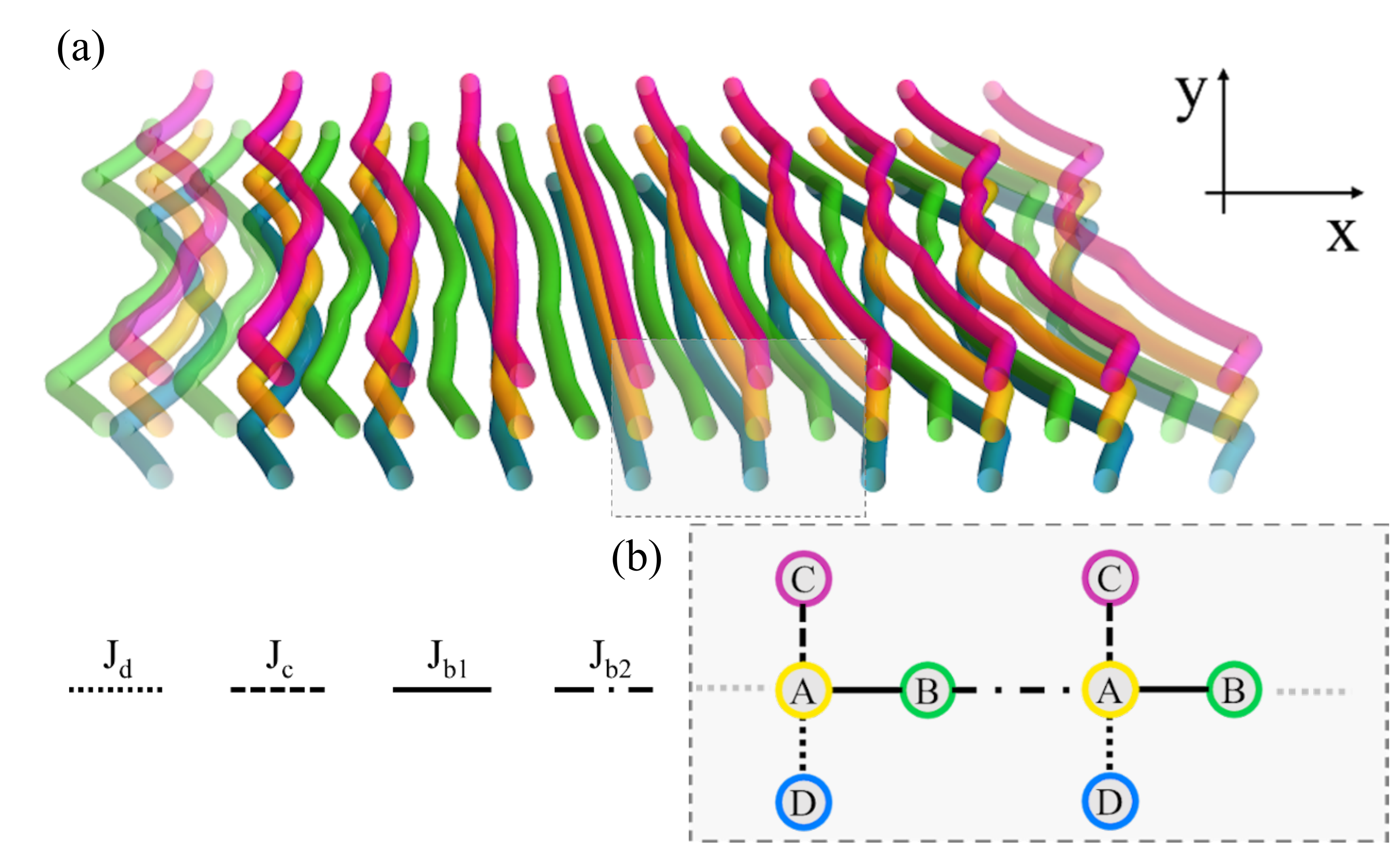}
		\caption{a) Illustration of the waveguides array with z-dependent couplings. b) Structure of the non-abelian waveguide lattice which features two dispersive and two degenerate flat bands.}\label{fig-structure}
	\end{center}
\end{figure}

The naL chain  consists of a one-dimensional lattice with  four sites per unit cell, indicated respectively as $A$, $B$, $C$ and $D$. As shown in Fig.\ref{fig-structure}, the lattice lies in the $(x,\,y)$-plane and it has two dangling bonds in each unit cell, it is an extension of the one-dimensional Lieb lattice considered in Refs.[\onlinecite{casteels2016,biondi2018}].
The inter- and intra- cell  hopping amplitudes along $x$ are indicated as $J_{b1}$ and $J_{b2}$ while $J_c$ and $J_d$ denote the hopping amplitudes along the dangling bonds. The Hamiltonian thus  reads as follows
\be
H_{\rm naL}\!=\!\!\!
%\sum_i  \lf[\k\, a^\dag_ia_i- \k\lf(b^\dag_ib_i+c^\dag_ic_i+d^\dag_id_i\rg)\rg] +\!\!\!\!\!\\
%& & \!\!\!\!\!\!\!\!\!\!\!\!+
\sum_i \!\lf(J_{b1}a^\dag_ib_i+J_{b2}a^\dag_{i}b_{i-1}+J_c a^\dag_ic_i+J_d a^\dag_id_i+ {\rm H.c.} \rg)
\ee
where $a_i,\, b_i,\,c_i,\,d_i$  and $a^\dag_i,\, b^\dag_i,\,c^\dag_i,\,d^\dag_i$ are bosonic annihilation and creation operators on the $A$, $B$, $C$ and $D$ sites of cell $i$. 
%The naL lattice, is a generalization of the one-dimensional Lieb lattice considered in Refs. \cite{casteels2016,biondi2018}.
%and  for $J_c=J_d=0$, it reduces to the Su-Schrieffer-Heeger(SSH) model \cite{su1979}.
%yields the Rice-Mele model \cite{rice1982} and if also $\k=0$ it 
Switching to $k$-space we can recast $H_{\rm naL}$ as follows 
\be H_{\rm naL}=\sum_k \!\lf(J_{b}a^\dag_kb_k+J_c a^\dag_kc_k+J_d a^\dag_kd_k+ {\rm H.c.} \rg)
 \ee
where $a_k,\,b_k,\,c_k$ and $d_k$
%$H_{\rm naL}=\sum_k \Phi^\dag_k {\cal H}_{k}  \Phi_k,$ where $\Phi^\dag_k$ and $\Phi_k=(a_k,b_k,c_k,d_k)$ 
denote $k$-space creation and annihilation operators and $J_{b}=J_{b1}+J_{b2} e^{ik}$.
%defined as $\phi_k=\sum_i \phi_i e^{i k R_i}$ with $R_i$ %=i\, a_{x}$, $a_x$ 
%indicating the position of cell $i$ while the matrix ${\cal H}_{k}$, with $J_{b}(k)=J_{b1}+J_{b2} e^{ik}$, has the following expression:
%\be\label{Hk}
%{\cal H}_{k}=
%\left(
%\begin{array}{cccc}
%	\kappa  & J_{b}(k)   & J_c & J_d   \\
%	J_{b}^*(k)  & -\kappa   &0&0   \\
%	J_c  &0   &-\kappa&0 \\
%	J_d &0&0&-\kappa 
%\end{array}
%\right)
%\ee
% 

%Most results presented here are 
%From the above equations one easily understands that photonic waveguide lattices are the ideal playground to implement the non-abelian Thouless pumping scheme proposed here.
%

Thouless pumping requires the cyclic modulation  of the parameters defining the lattice potential.  %and of the longitudinal wavevector $\kappa$.  
%In its abelian version, it has been realized with ultracold atoms by  dynamically controlling the optical superlattice \cite{lohse2015,nakajima2016} and in modulated photonic waveguide lattices \cite{kraus2012,wang2019}  where it leads to topologically quantized light propagation.
%%
%For this purpose different protocols have been developed for different systems ranging from cold atoms \cite{lohse2015,nakajima2016} to photonic waveguide lattice
%Here we focus on photo
%Here we focus on photonic waveguide lattices, but most results can be extended to ultracold atoms setups.
%
% Thouless pumping can be implemented  by modulating the tunneling amplitudes and the longitudinal wavevector along $z$.
Clearly, rather than a charge current, photonic lattice pumping  yields a $z$-dependent displacement of the photon beam across the lattice \cite{ke2016,wimmer2017,fedorova2019,wang2019}.  %Topological quantization then leads to topologically quantized light propagation. 
Specifically, assuming that the system is initialized in a Wannier state, the role of the pumped charge is  played by the displacement of the Wannier center, whose geometric and topological significance,  was first elucidated in the context of  polarization theory \cite{resta2007}.
Below we establish a general relation between displacement and non-abelian holonomies.

%From the above equation one can easily understand the analogies and differences between  the standard 2D Lieb lattice \cite{lieb1989} and the naL chain.%: first the one-dimensional character and second the presence of a larger unit cell

 \section{Displacement and holonomies}
 \label{sect-main}

As schematically shown in Fig.\ref{fig-structure}a),  we assume that the hopping amplitudes, $J_\mu$ with $\mu=b1,b2,c,d$ are periodic functions of the longitudinal coordinate $z$ and we indicate with $\lambda_0$ the wavelength of their modulation.

To calculate the photon beam displacement, $\Delta x$, generated along a cycle we start from the following general expression %\cite{sup}
\be\label{deltax-general}
\Delta x=\int_{0}^{\lambda_0}\partial_z (\la\psi(z)|\hat x|\psi(z)\ra) dz
\ee
where $|\psi(z)\ra$ indicates  the electromagnetic field's amplitude, solution of a Schr\"odinger equation of the form \cite{bra-ket}, 
\be\label{eq:sch}
i\partial_z |\psi(z)\ra= H_{\rm naL}(z)|\psi(z)\ra
\ee and  $\hat x$ denotes the position operator.
%, {\sl i.e.}
%${\hat x=\sum_{k}\!  |m_k\ra\la m_k | i\partial_k}$, with the index $m$ enumerating the sites within a unit cell, {\sl i.e.} $m \in [a,b,c,d]$. 
%To arrive at an analytic expression of the relation between displacement and holonomies, we focus on the {\sl adiabatic limit} \cite{messiah1999,sup};  we thus assume
 Assuming that $\lambda_0$ is the largest relevant length-scale, in the adiabatic limit, a suitable basis to expand the field $|\psi(z)\ra$ is the Bloch eigenmodes basis, $\{|\psi_{\nu a}(k,z)\ra\}$,  that diagonalizes the naL Hamiltonian for each value of $k$ and $z$. See Appendix \ref{spectrum} for more details.
%\be
%{\cal H}_k(z)|\psi_{\nu \a}(k,z)\ra=\kappa_\nu(z)|\psi_{\nu \a}(k,z)\ra.
%\ee
%
%%Having in mind the photonic waveguide lattice implementation,and the wavevector $\k$ are independent periodic functions of the longitudinal coordinate $z$ and indicate with $\lambda_0$ the wavelength of their modulation. We require $\lambda_0$ to be the largest relevant length-scale in the problem so that the adiabatic theorem holds. 
%For each value of $z$ and $k$,  the set $\{|\psi_{\nu \a}(k,z)\ra\}$
It consists of  two dispersive modes % %
% \be
%  \psi_{\pm}=\tilde N_{\pm}\lf[(\sqrt{|J|^2+\k^2} \pm \k)|k,A\ra\pm \lf(J(k)|k,B\ra+J_c|k,C\ra+J_d|k,D\ra\rg)\rg]
% \ee
 with longitudinal momenta $\kappa_{\pm}(k)=\pm\Delta(k)$, $\Delta(k)=\sqrt{|J_b(k)|^2+J_c^2+J_d^2}$  and 
two dispersionless %(CONTROLLA TERMINI)%
degenerate modes, $|\psi_{01} \ra$ and $|\psi_{02}\ra$, 
  with longitudinal momentum $\kappa_{0}=0$, defined as
\bea
|\psi_{01}\ra &=&\frac{J_c |d_k\ra-J_d|c_k\ra}{\delta} \label{eq:basisNA1}
\\
| \psi_{02}\ra &= &\frac{\delta^2 |b_k\ra+J_{b}^*(J_c|c_k\ra+J_d|d_k\ra)}{\delta\Delta(k)} \label{eq:basisNA2}
\eea
with $\delta=\sqrt{J_c^2+J_d^2}$.

%In what follows we only need the explicit expression of the degenerate Bloch eigenmodes,   
To keep the discussion general, we assume that  the system is initialized, at $z=z_0,$ in the Wannier state centered at the site $n$ belonging to the mode $\kappa_\nu$,
we thus set
\be\label{eq:in-cond}
|\psi(z_0)\ra=|w_{\nu a}(n,z_0)\ra= \sum_{ka}c_a|\psi_{\nu a}(k,z_0)\ra e^{ikn}
\ee
where $k$ runs over the reciprocal lattice sites, $\nu=0,\pm$ and  the subscript $a$, not to be confused with the operator $a$, enumerates the basis states within the degenerate subspace, {\sl i.e.} $a\in[1,\ldots,d_\nu]$ with $d_\nu$ indicating the dimension of the degenerate modes subspace.
For the naL lattice we have $d_0=2$ and $d_{+}=d_-=1$.
%Using the initial condition of  Eq.(\ref{eq:in-cond})  we arrive at the following expression for the field at  $|\psi(z_1)\ra$ in the adiabatic limit:
%%
%\be\label{W1NA}
%|\psi(z_1)\ra=\sum_{kab} c_a \lf[W_\nu(z_1,z_0)\rg]_{ba} |\psi_{\nu b}(k,z_1)\ra e^{ikn} %=\sum_{k} e^{ikn}  \lf[|\Psi_{\nu}(k,z_1)\ra W_\nu(z_1,z_0)\rg]_a
%\ee
%%
%where  $W_\nu(z_1,z_0)$ can be written as the product of a dispersive term and  a geometric contribution as follows %\eqref{eq-W}. 
%\be\label{W2NA}
%W_\nu(z,z_0)=e^{i\!\int_{z_0}^z \!\k_{\nu}(z)dz}\,{\cal P} \exp \lf[i\!\int_{z_0}^z\!\Gamma^z_\nu dz\rg]\ee
%%
%with $\Gamma^z_{\nu }$ denoting the non-abelian connection of Wilczek and Zee \cite{wilczek1984}  $\lf[\Gamma^z_{\nu }\rg]_{ab} = \la\psi_{\nu a }|i\partial_z| \psi_{\nu b }\ra$ and $\cal P$ the path-ordering.
%%When  $\k_\nu$ corresponds to a non-degenerate mode $\Gamma^z_{\nu }$ reduces to the standard Berry's connection \cite{wilczek}.
%%\\
%%\\

Since the modulation of the lattice parameters preserves the translational invariance of the lattice  for all $z$, the quasi-momentum, $k$, is a good quantum number both in the driven and undriven system and 
the adiabatic evolution operator, $U(z, z_0)$, can be factorized as follows
 \be U(z, z_0)\simeq\prod_{\mu k} {\cal U}_{\mu k}(z, z_0).\label{AD-EV1}\ee
In the above equation ${\cal U}_{\mu k}(z, z_0)$ describes the evolution starting from a local Bloch eigenmode  with quasi-momentum $k$ and longitudinal  momentum $\k_\mu$,
% When $\k_\mu$ corresponds to the set of degenerate modes, ${\cal U}_{\mu k}(z, z_0)$ is given by:
 %
 \be {\cal U}_{\mu k}(z, z_0)=\sum_{ab}\lf[W_{\mu}\rg]_{ab}(z,z_0) |\psi_{\mu a}(k,z)\ra\la \psi_{\mu b}(k,z_0)|\label{AD-EV-BLOCH-NA}\ee
where  the symbol
$[\dots]_{ab}$ denotes the
$ab$ element of the matrix inside the brackets, $a$ and $b$ span a degenerate subset and $W_{\mu}(z,z_0)$ is \cite{wilczek1984}:
\be W_\mu(z,z_0)=e^{i\!\int_{z_0}^z \!\k_{\mu}(z)dz}\,{\cal P} \exp \lf[i\!\int_{z_0}^z\!\Gamma^z_\mu dz\rg]\label{eq-W} \ee
with ${\cal P}$ denoting the path-ordering and  $\Gamma^z_\mu$ indicating the Wilczek-Zee connection,
 \be\lf[\Gamma^z_{\mu}\rg]_{ab}= \la\psi_{\mu a}(k,z)|i\partial_z| \psi_{\mu b}(k,z)\ra\label{gamma}.\ee
 
When $\k_\mu$ corresponds to a single non-degenerate mode, $\Gamma^z_{\mu}$ reduces to the standard Berry's connection.

Using the adiabatic evolution operator defined in equations (\ref{AD-EV1}-\ref{gamma}) and the initial condition given in  Eq.(\ref{eq:in-cond}),  we obtain the following expression for the field at $z=z_1> z_0$ in the adiabatic limit   
\be\label{psiz1}
|\psi(z_1)\ra=\sum_{kab} c_a \lf[W_\nu(z_1,z_0)\rg]_{ba} |\psi_{\nu b}(k,z_1)\ra e^{ikn}. %=\sum_{k} e^{ikn}  \lf[|\Psi_{\nu}(k,z_1)\ra W_\nu(z_1,z_0)\rg]_a
\ee

From the above equation, it correctly emerges that the adiabatic dynamics in the presence of $N$ degenerate modes is invariant under $z$-dependent  changes of basis in the degenerate subset, {i.e.} $SU(N)$ gauge transformations. Consequently,  $W$ transforms as:
$$ W(z,z_0) \rightarrow M^{\dag}(z) W(z,z_0) M(z_0)$$
and for cyclic transformations it behaves as a rank 2 tensor.  It yields, apart from an irrelevant phase factor, the Wilson loop, $W(\lambda_0,0)$, on the fiber bundle that is locally the product of the $z$-varying parameters space and the degenerate modes subset. %In particular, if N=1, $W(z,z_0)$ reduces to the Berry's phase factor.
%%
%where  $W_\nu(z_1,z_0)$ can be written as the product of a dispersive term and  a geometric contribution as follows %\eqref{eq-W}. 
%\be\label{Wzz0NA}
%W_\nu(z,z_0)=e^{i\!\int_{z_0}^z \!\k_{\nu}(z)dz}\,{\cal P} \exp \lf[i\!\int_{z_0}^z\!\Gamma^z_\nu dz\rg]\ee
%%
%with $\Gamma^z_{\nu }$ denoting the non-abelian connection of Wilczek and Zee \cite{wilczek1984}  $\lf[\Gamma^z_{\nu }\rg]_{ab} = \la\psi_{\nu a }|i\partial_z| \psi_{\nu b }\ra$ and $\cal P$ the path-ordering.
%%When  $\k_\nu$ corresponds to a non-degenerate mode $\Gamma^z_{\nu }$ reduces to the standard Berry's connection \cite{wilczek}.
%%\\
%%\\

Replacing Eq.(\ref{psiz1}) in Eq.\eqref{deltax-general}, we can derive a transparent and simple expression for the displacement introducing the non-abelian connection along $k$, {\sl i.e.} $\lf[\Gamma^k_{\nu }\rg]_{ab} = \la\psi_{\nu a }|i\partial_k| \psi_{\nu b }\ra$, that is the non-abelian version of the Zak phase \cite{zak1989}. 
By doing so,  following the derivation described below, we arrive at our final expression for the photon beam displacement:
\be\label{eq:deltaxNA1}
 \Delta x=\sum_{ab}c^*_a c_b D^{\nu}_{ab}\, ,
\ee
 where the displacement matrix $D^{\nu}_{ab}$ can be expressed as follows
\be\label{eq:displ-matr}D^{\nu}_{ab}=\frac{1}{2\pi} \int_{0}^{\lambda_0}\!\!\!dz \int_{-\pi}^{ \pi}  \!\! \!\!  dk \lf[
W^\dag_\nu\,{\cal F}^\nu_{kz}W_\nu \rg]_{ab} \ee
with ${\cal F}^\nu_{kz}=\partial_k \Gamma^z_\nu-\partial_z \Gamma^k_\nu+i\lf[\Gamma^z_\nu, \Gamma^k_\nu\rg]$ denoting the non-abelian field strength  matrix. %in vector notation, i.e. $|\Psi_{\nu}\ra=(|\psi_{\nu1}\ra,..., |\psi_{\nu d_\nu}\ra )$ where $d_\nu$ is the dimension of the degenerate subspace, 

Equations (\ref{eq:deltaxNA1}-\ref{eq:displ-matr}) relate the photon beam displacement to the non-abelian holonomy generated along a cycle when the evolution involves a degenerate eigenmodes subspace. It clearly shows that in general, the displacement, $\Delta x$, will bear the consequence of the non-abelian nature of the dynamics while it yields the known abelian result when $d_\nu=1$.
Comparing  Eqs. (\ref{eq:deltaxNA1}-\ref{eq:displ-matr}) to their abelian counterpart \cite{resta2007,thouless1983}, one easily realizes that the displacement matrix  can be viewed as the flux of the non-abelian field strength, ${\cal F}^\nu_{kz}$, that transforms along the path with the path-dependent factors $W^\dag_\nu$ and $W_\nu$ as prescribed by the non-abelian Stokes theorem \cite{fishbane1981}.
As shown in Ref. [\onlinecite{fishbane1981}], the presence of these factors ensures that the surface integral is independent on the choice of the surface. Furthermore it guarantees that the displacement matrix, $D^{\nu}_{ab}$, transforms as a rank-2 tensor under gauge transformation and consequently that the total displacement given by Eq. \eqref{eq:deltaxNA1} is gauge-invariant.

%, transforms along the path.
Before proceeding further, we present a derivation of Eqs.  (\ref{eq:deltaxNA1}-\ref{eq:displ-matr}).
To start with we express  the position operator $\hat x$ in  $k$-space as follows,
%$\hat x$ can be recast as
\be \label{xop}
\hat x=\sum_{m,k}  |m_k\ra\la m_k| i\partial_k,
\ee
where $m=a,b,c,d$. %$ |m_k\ra$ are the standard Bloch vectors. %defined in Appedix \ref{app1}.
%enumerating the 
Inserting  \eqref{xop} in Eq.\eqref{deltax-general}  and using Eq.\eqref{psiz1}  and \eqref{eq:deltaxNA1} we can rewrite the displacement matrix as follows
\bea
D^\nu_{ ab}&=&\frac{1}{2\pi}\int_{0}^{\lambda_0}\int_0^{2 \pi} \!\!  \partial_z \lf[ W^\dag_\nu(z,z_0)\la\Psi_\nu(k,z)| \rg.\nn\\& & \lf.  i\partial_k\lf(|\Psi_{\nu}(k,z)\ra W_\nu(z,z_0)\rg)\rg]_{ab}\, dk\,dz\nn\\ \label{dim1}
\eea
% the subscript ${\nu a}$ recalls  the initial condition and
where we introduced the vector notation, i.e. $|\Psi_{\nu}\ra=(|\psi_{\nu1}\ra,..., |\psi_{\nu d_\nu}\ra )$.
Expanding the derivatives in equation \eqref{dim1} we obtain:
\bea
D^\nu_{ ab}&=&\frac{1}{2\pi}\int_{0}^{\lambda_0}\!\!\!dz\int_0^{2 \pi}\!\! \!\!dk  \lf[\partial_z \lf(W_\nu^\dag\lf<\Psi_{
\nu}\rg|\,i\partial_k\,\lf|\Psi_{\nu}\rg>W_\nu\rg)+
\rg.\nn\\ &&\lf.+i\partial_z\lf(W_\nu^\dag\rg)\,\partial_k\,W_\nu+iW_\nu^\dag\,\partial_k\partial_z\,W_\nu
\rg]_{ab}\nn\\
\label{dim2}\eea
 The $z$-derivatives of the operators $W_\nu$ and $W^\dag_\nu$ are given by:
 \bea \label{WEQ1}\partial_zW_\nu&=&i(\Gamma_\nu^z+\k_{\mu}I)\,W_\nu\\
 \partial_z
W^\dag_\nu&=&-iW_\nu^\dag\,(\Gamma_\nu^z+\k_{\mu}I) \label{WEQ2} \eea
where $\Gamma_\nu^z$ is the non-abelian connection defined in Eq.\eqref{gamma}.
Substituting these relations in Eq.(\ref{dim2}) we obtain:

\bea\label{dim3}D^\nu_{ ab}&=&\frac{1}{2\pi}\int_{0}^{\lambda_0}\!\!\!dz\int_{-\pi}^{\pi} \!\! \!\! dk \lf[-
W^\dag_\nu\,\partial_k\lf(\Gamma_\nu^z+\k_{\mu}I\rg)W_\nu+\rg.\nn\\& & \lf. -\partial_z\lf(W_\nu^\dag \Gamma_k W_\nu\rg)\rg]_{ab}.\eea

Using again Eqs.(\ref{WEQ1}-\ref{WEQ2})  to express the $z$-derivatives and using the definition of ${\cal F}_{zk}$ and the $k$-periodicity of $\kappa_\mu$,
we easily recover Eq. \eqref{eq:displ-matr}.

\section{Pumping cycles}
\label{sect-cycles}
\begin{figure}[t]
	\begin{center}
		\includegraphics[width=\columnwidth]{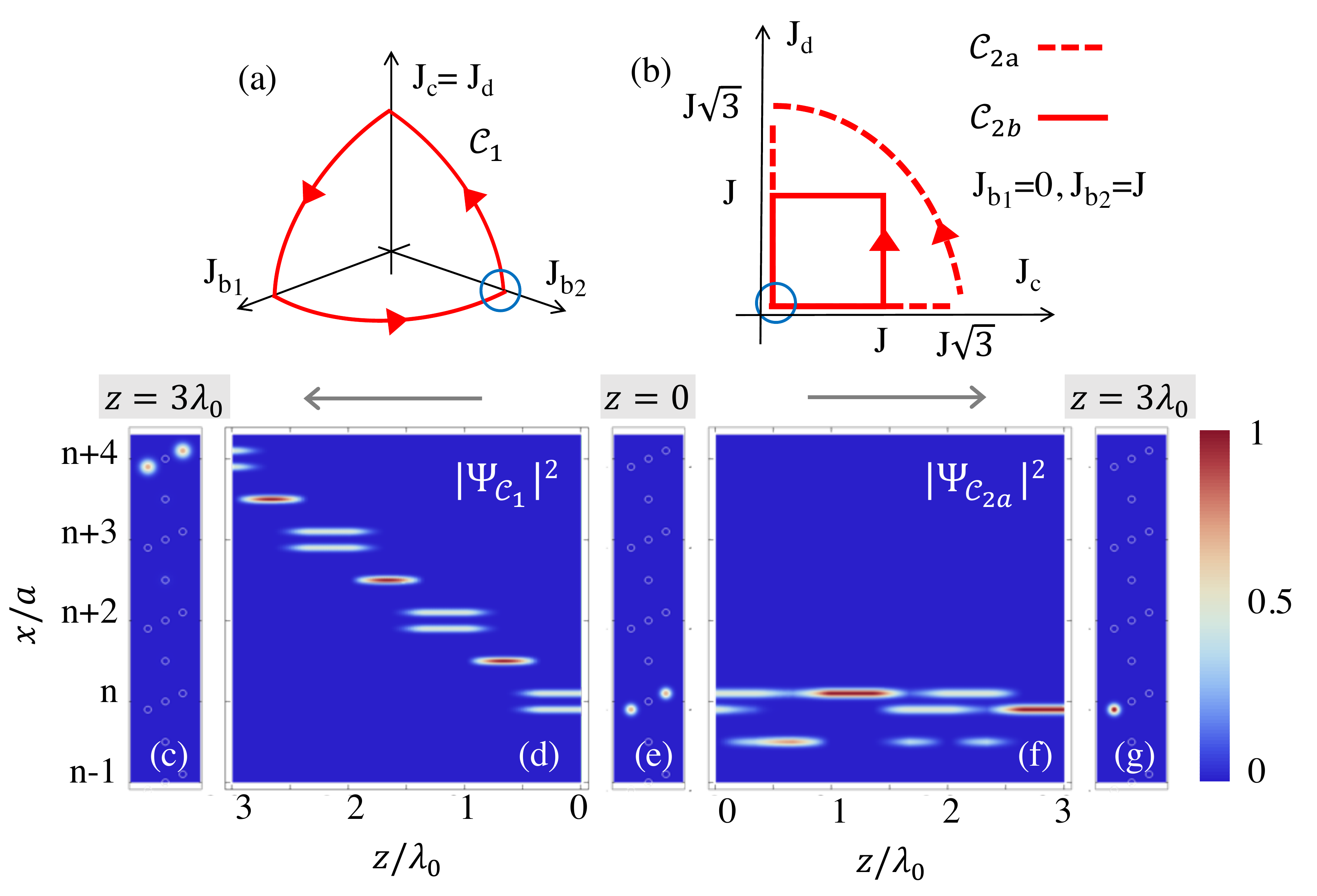}\\
		\caption{(a-b) Examples of pumping cycles. The blue circle indicates their starting point, $z=z_0$. Note that we take a small but finite values of $J_c=J_d$ at $z=z_0$ to avoid ambiguities in the definition of the eigenstates.   Along ${\cal C}_1$ we have $J_d=J_c$ while along ${\cal C}_{2(a-b)}$  we have $J_{ b2}=J, J_{ b1}=0$. (c-g) Numerically evaluated field's intensity along ${\cal C}_1$ and ${\cal C}_{2a}$ with $\lambda_0J=200$, panels (c) and (g) show the field intensity in the final states after three cycles $C_1$ and  after three cycles $C_{2a}$. To allow a clearer identification, in all panels we slightly displaced the sites ``C'' and ``D'' along $x$. }\label{noncom}
	\end{center}
\end{figure}

To  further understand the non-commutative nature of pumping in the naL chain, it is useful to consider some specific example. In Fig. \ref{noncom}(a-b) we show three pumping cycles, ${\cal C}_1$,  ${\cal C}_{2a}$ and ${\cal C}_{2b}$.
As one can see, the cycle ${\cal C}_1$ defines a spherical triangle in parameters space  while the cycles ${\cal C}_{2a}$ and ${\cal C}_{2b}$ 
are  contained in the plane $J_{b2}=J$ and $J_{b1}=0$ and they involve only the manipulation of $J_c$ and $J_d$.
The blue circle indicates the point where we start covering the cycles at $z=z_0$.  
The  corresponding holonomy transformations can be written as, (see Appendix \ref{app-cycles} for more details)
\be \label{holonomies}
W_{{{\cal C}_1}}=e^{i\frac{k}{2}\lf(\sigma_0-\sigma_z\rg)}\,\, {\rm and} \,\, W_{{\cal C}_{2i}}=e^{i\Delta_{{\cal C}_{2i}}\!\lf( \sin(k)\sigma_x+\cos(k)\sigma_y\rg)}%$ with $\Delta_{{\cal C}_{2a}}=\pi/4$ and $\Delta_{{\cal C}_{2b}}=\pi/6$.
\ee
where $\sigma_x,\,\sigma_y,\,\sigma_z\,\,{\rm and }\,\,\sigma_0$ denote respectively the Pauli matrices and the $2\times2$ identity matrix and the angles $\Delta_{{\cal C}_{2i}}$ depend on the precise shape of the cycles ${\cal C}_{2i}$. The above expressions hold in the basis of  Bloch eigenmodes at $z=z_0$.  Using
Eqs. (\ref{eq:basisNA1}-\ref{eq:basisNA2}),  the latter can be shown to satisfy the following simple relations: $|\psi_{01} \ra \propto |d_k\ra-|c_k\ra$ and  $|\psi_{02} \ra \propto |d_k\ra+|c_k\ra.$
%At this point the degenerate modes can be expressed as and localized on the cell $n$ can be simply expressed as: 
%\be |\psi_{01}(z_0,k)\ra=\frac{|c_k\ra-|d_k\ra}{\sqrt{2}} \,\, {\rm and} \,\, |\psi_{02}(z_0,k)\ra=e^{ik}\frac{|c_k\ra+|d_k\ra}{\sqrt{2}}\ee
%
%
From Eq. \eqref{holonomies} we thus see, that in the adiabatic limit,  the cycle  ${\cal C}_1$ does not affect the antisymmetric mode, $|\psi_{01}\ra$, while it simply multiplies the symmetric mode,  $|\psi_{02}\ra$, times the phase factor $e^{ik}$. The cycles ${\cal C}_{2a}$ and ${\cal C}_{2b}$ instead yield a rotation in the degenerate subspace by an angle $\Delta_{{\cal C}_{2i}}$ such that
 $\Delta_{{\cal C}_{2a}}=\pi/4$  and $\Delta_{{\cal C}_{2b}}=\pi/6$, as shown in Appendix \ref{app-cycles}.

These approximate analytical results are in good agreement with the numerical results, shown in Figures \ref{noncom}(c-g), obtained by solving  Eq.\eqref{eq:sch} on a finite lattice.
%obtained solving by solving numerically  Eq.\eqref{eq:sch}.
%{\sl i.e.} (see \cite{sup} for more details)
% %The above expression 
%The above matrices, $W_{{{\cal C}_1}}$ and $W_{{\cal C}_{2i}}$, represent  the holonomy transformations in the Bloch eigenmodes basis, Eqs. (\ref{eq:basisNA1}-\ref{eq:basisNA2}), at $z=z_0$.
% and, as stated by Eq.\eqref{W1NA},  they define the evolution in the adiabatic limit. %as it can be shown by a simple calculation \cite{sup},   
%defined as
%\be|\psi_{01}(z=z_0)\ra=(|c_k\ra+|d_k\ra)1/{\sqrt 2}\ee
%By looking at hey lead to results in good agreement  with those shown in 
There we plot the field's intensity as a function of $z$ and $x$ during the application of three consecutive cycles ${\cal C}_1$, Fig.\ref{noncom}(d), and in the final state at $z=3\lambda_0$, Fig.\ref{noncom}(c). Analogous plots for the cycle ${\cal C}_{2a}$ are shown in Fig.\ref{noncom}(f-g). In both cases the system is initialized in the Wannier state $|w_{02}(n, z_0)\ra$, corresponding to the symmetric mode, as shown Fig. \ref{noncom}(e) displaying the field's intensity  at $z=z_0$.
%As expected from Eqs. \eqref{holonomies}, 
Comparing Fig. \ref{noncom}(c) and  Fig. \ref{noncom}(e) we see that  the application of three cycles ${\cal C}_1$  displaces the initial state forward by three unit cells. On the contrary  the application of three cycles ${\cal C}_{2a}$ yields a $3/4\pi$ rotation but zero displacement, thus the initial state $|c_n\ra+|d_n\ra$ is rotated into the state $|c_n\ra$ as shown in  Fig. \ref{noncom}(g). %From these Figures  it emerges quite clearly that,  while the cycle ${\cal C}_{1}$ simply displaces  the state $|w_{02}(n,z_0)\ra$ by one unit cell, 
%the ${\cal C}_{2a)}$ simply rotates the initial state by $\pi/4$ yielding a field localized in on the waveguide $c$ of the cell $n$. %since along these cycles the coupling $J_{b1}$ equals zero and the lattice reduces to a chain of disconnected dimers.
\begin{figure}[t]
	\begin{center}
	\includegraphics[width=1\columnwidth]{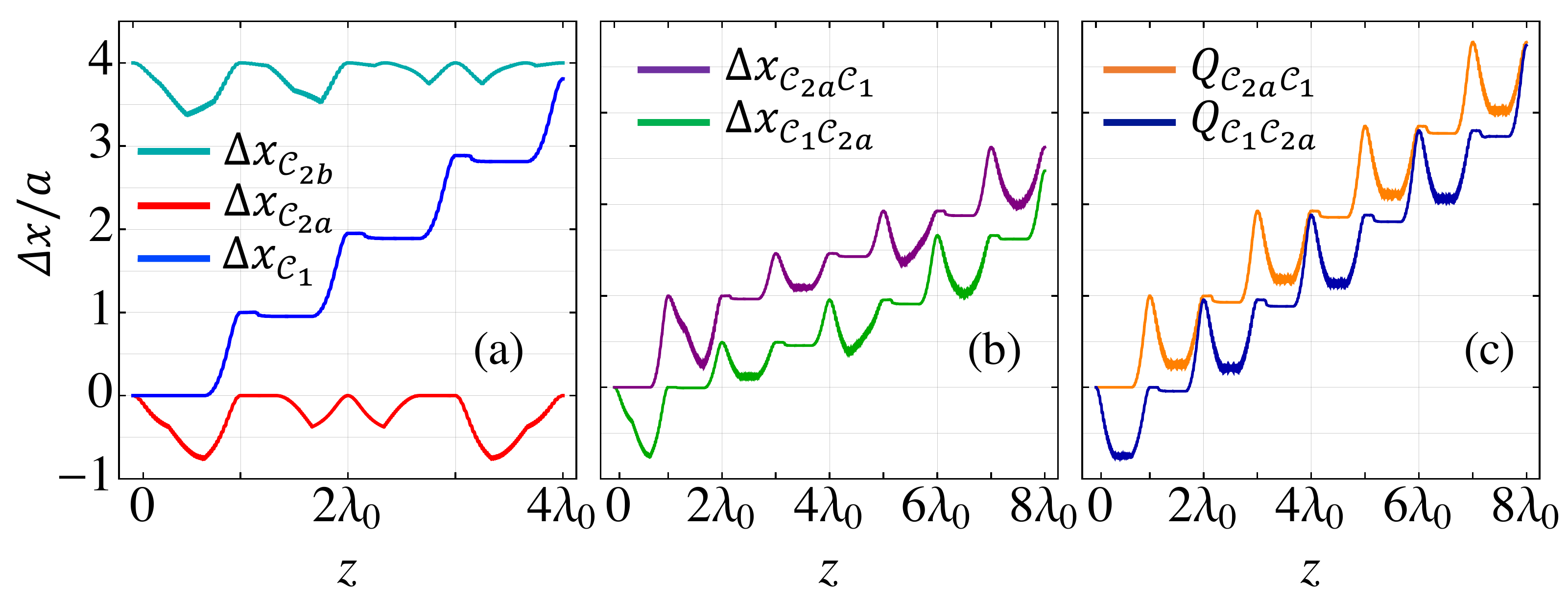} 
		\caption{ Photon beam displacement for a) the cycles ${\cal C}_{1}$ (middle line), ${\cal C}_{2a}$ (lower line) and ${\cal C}_{2b}$ (upper line)  and b) the sequential application of ${\cal C}_{1}$ and ${\cal C}_{2a}$:  ${\cal C}_{2a}{\cal C}_{1}$ (upper line)  and ${\cal C}_{1}{\cal C}_{2a}$ (lower line). c) Trace of the displacement matrix for the sequences ${\cal C}_{2a} {\cal C}_1$ (upper line)  and ${\cal C}_1 {\cal C}_{2a}$ (lower line). }\label{displ}
	\end{center}
\end{figure}

The difference between the cycles  ${\cal C}_{1}$, ${\cal C}_{2a}$ and ${\cal C}_{2b}$ can be also appreciated in Fig. \ref{displ}. There we plot the displacement evaluated numerically from equation  \eqref{deltax-general} for the system initially prepared in the state   $|w_{02}(n,z_0)\ra$ as a function of $z$ for the different cycles.
In panel (a) we see that the displacement generated along the cycle ${\cal C}_{1}$  is  quantized. This is a natural consequence of the structure of the holonomy matrix,  given in Eq. \eqref{holonomies},  that leads to the following simple expression for  the displacement matrix  $D^{0}_{{\cal C}_1}=
\left(
\begin{array}{cc}
0  & 0    \\
 0 & 1     \\  
\end{array}
\right), 
$ as discussed in details in Appendix \ref{app-cycles}.
%
%
%The cycle does not induce any mixing between the two states since the associated holonomy transformation given in Eq. \eqref{holonomies} is diagonal. 

This quantization  reflects the topological nature of Thouless pumping that in the non-abelian case implies,
$
{\rm Tr}[ D^{\nu}_{\cal C}]=C^{1}_{\nu}
$
where $C^{1}_{\nu}$ denotes the  first Chern number of the degenerate band $\nu$ \cite{manton2004}.

% for the definition of the Chern number in the non-abelian case.  % and  denotes the corresponding first Chern number.
%Note that depending on the cycle we may have  different values of  ${\rm Tr}[ D^{\nu}]$
%
%
%The non-abelian nature of the evolution and its topological properties are intrinsically related.
%The integer unit quantization of the displacement generated by cycle ${\cal C}_1$  emerging in Figure \ref{displ}(a) can be explained considering that along this cycle the two degenerate modes
% are not mixed and the displacement matrix  for the degenerate modes band has the simple structureTherefore,  if the system is initially prepared in the state  $|w_{02}(n,z_0)\ra$, a unitary displacement will be generated. %and the evolution is adiabatic, 
As shown in Fig. \ref{displ}(b), the situation is rather different if we perform a sequence of the cycles ${\cal C}_1$ and ${\cal C}_{2a}$. In this case the non-abelian nature of the evolution manifests in a non-integer displacement per cycle and in the dependence of the generated displacement on the ordering of the sequence. Focusing on the sequence  ${\cal C}_1{\cal C}_{2a}$  we see that starting from the state $|w_{02}(n,z_0)\ra$ we get a unitary displacement;  for the other ordering, first ${\cal C}_{2a}$ and then ${\cal C}_1$, the displacement equals 1/2.
The perfect quantization can be recovered by considering the trace of the displacement matrix\cite{Qianqian2020,Cerjan2020}, {\sl i.e.} the sum of the displacement generated starting in the state  $|w_{02}(n,z_0)\ra$ and  $|w_{01}(n,z_0)\ra$. This is exactly what can be seen in Fig. \ref{displ}(c) where we plot the evolution  as a function of $z$ of the traces of the displacement matrices for the sequences  ${\cal C}_{2a} {\cal C}_1$ and ${\cal C}_1 {\cal C}_{2a}$, denoted as $Q_{{\cal C}_{2a} {\cal C}_1}$ and  $Q_{{\cal C}_{1} {\cal C}_{2a}}$.

Non-commutative effects  can be detected in various ways in the evolution of the field. For example, one could design a complex pumping cycle aimed at detecting the contribution of the commutator term in the non-abelian Berry’s curvature and consequently the difference between the abelian and non-abelian prediction for the pumped charge. Here, we follow an alternative route, namely we consider the simple cycles ${\cal C}_{2i}$ and ${\cal C}_1$  and we show how the non-abelian nature of the dynamics may be detected by changing the order of the cycles.

\begin{figure}[t]
	\begin{center}
			\includegraphics[width=\columnwidth]{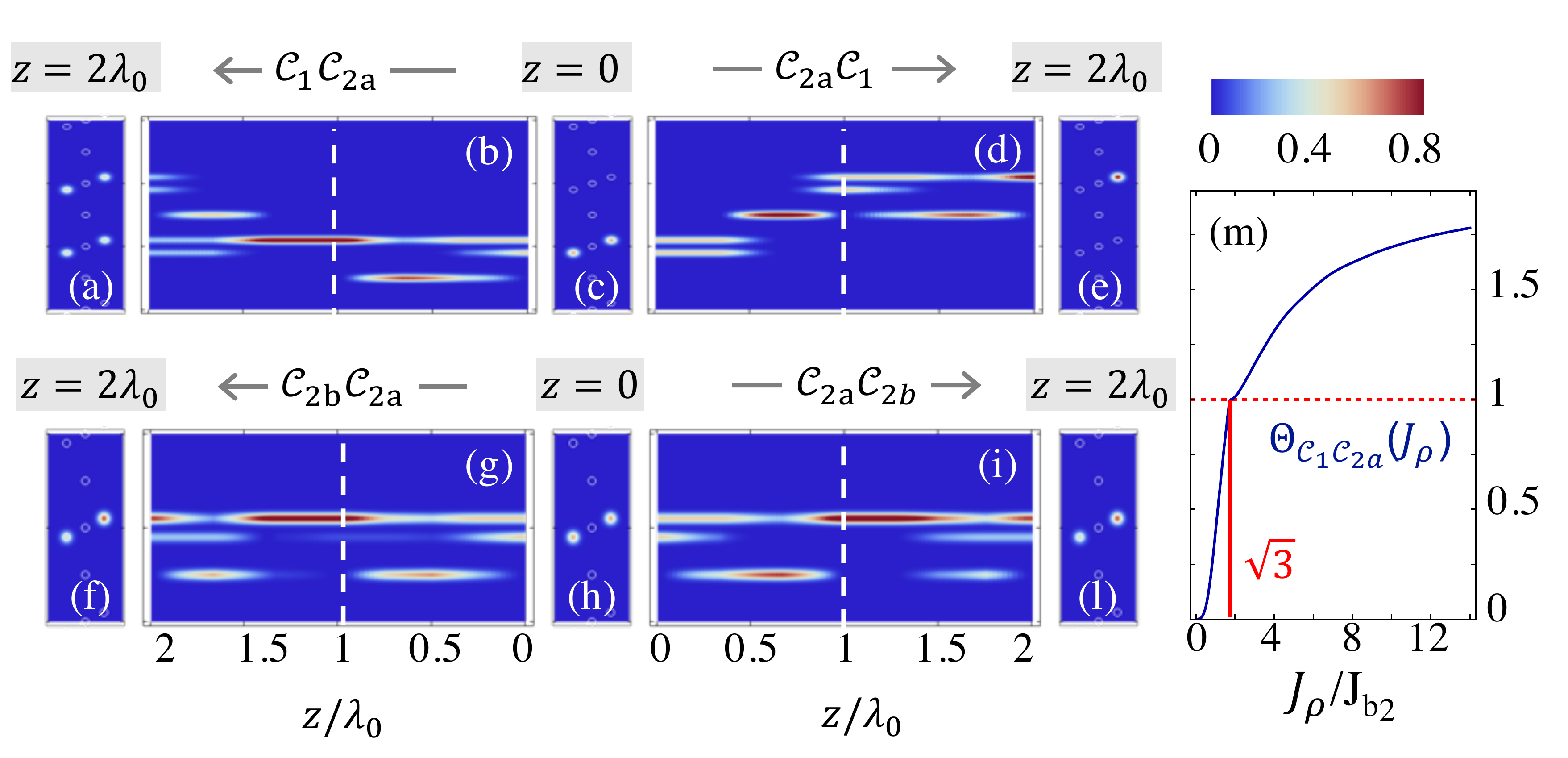}
%		\hspace{0.1cm}
%		\includegraphics[width=0.4\columnwidth]{Fig2-2.pdf}\\ 
%		\vspace{0.2cm}
%		\includegraphics[width=0.4\columnwidth]{Fig2-3.pdf} 
%		\hspace{0.1cm}
%		\includegraphics[width=0.4\columnwidth]{Fig2-4.pdf} 
		\caption{(a-l) Numerically evaluated field's intensity along different cycles sequences involving  ${\cal C}_{1}$, ${\cal C}_{2a}$ and ${\cal C}_{2b}$. For non-commuting cycles the final state depends on the ordering of the sequence. Parameters as in Fig. \ref{noncom}. (m) \textit{Distance} between the final states obtained with ${\cal C}_{1}{\cal C}_{2a}$ and with ${\cal C}_{2a}{\cal C}_{1}$. The evolution along the sequence  ${\cal C}_{2a}{\cal C}_{1}$, i.e. from panel (c)  to panel (e) is shown dynamically in the video NAevolution.mp4. } %Panels (a) and (b) show the non-commutativity of cycles  ${\cal C}_{1}$ and ${\cal C}_{2a}$. }
		\label{non-ab}
	\end{center}
\end{figure}

 To this end,  in  Fig.\ref{non-ab}(a-l)  we plot the field intensities as a function of $z$ for different cycles sequences and different orderings.  As expected, we see that, when the cycles generate non-commuting holonomies, the final state depends on the ordering of the cycles,  as shown in Figs. \ref{non-ab}(a-e) for the cycles ${\cal C}_{1}$ and ${\cal C}_{2a}$.
Specifically, we see that performing the sequence  ${\cal C}_{2a}{\cal C}_{1}$, the initial state, proportional to $|c_n\ra+|d_n\ra$, is first displaced and subsequently rotated by $\pi/4$ yielding the final state $|c_{n+1}\ra$, Fig. \ref{non-ab}(d-e). In the opposite case, ${\cal C}_{1}{\cal C}_{2a}$, the state is first $\pi/4$-rotated and subsequently displaced through the  cycle ${\cal C}_{1}$, the final state is thus a linear combination of $|c_n\ra-|d_n\ra$ and $|c_{n+1}\ra+|d_{n+1}\ra$.
On the contrary, when the holonomies of the two cycles commute, the final state does not depend on the ordering of the cycles, as shown for the cycles ${\cal C}_{2a}$ and ${\cal C}_{2b}$ in Figs. \ref{non-ab}(f-l).

A figure of merit, $\Theta$, to characterize  the efficiency in generating non-abelian effects may be thus defined from the \textit{distance} between the final states obtained performing the sequence of two cycles in opposite orders, {\sl i.e.}
$\Theta_{{{\cal C}_1{\cal C}_{2i}}}=\sum_{n,m}\lf(|\la m_n|\psi_{{\cal C}_1{\cal C}_{2i}}\ra|-|\la m_n|\psi_{{\cal C}_{2i}{\cal C}_1}\ra|\rg)^2$ where $m$ and $n$ enumerate respectively the sites in a unit cell  and the unit cells in the lattice.
The value of $\Theta$ will be affected in particular by the shape of the cycle and by the initial conditions. In Figure \ref{non-ab}(m) we show the dependence of  $\Theta_{{\cal C}_1{\cal C}_{2a}}$ on the maximum value, $J_{\rho}$, acquired by $J_c$ and $J_d$ in the cycle  ${\cal C}_{2a}$, that was previously set to $J_{b2}\sqrt{3}$ with $J_{b2}=J$. We see that $\Theta$ increases as we increase $J_{\rho}$ tending towards its maximum value equal to 2. The latter is reached asymptotically when $C_{2a}$ yields a rotation of $\pi/2$. Further discussion of these effects can be found in Appendix \ref{app-cycles}.

\begin{figure}[t]
	\begin{center}
			\includegraphics[width=\columnwidth]{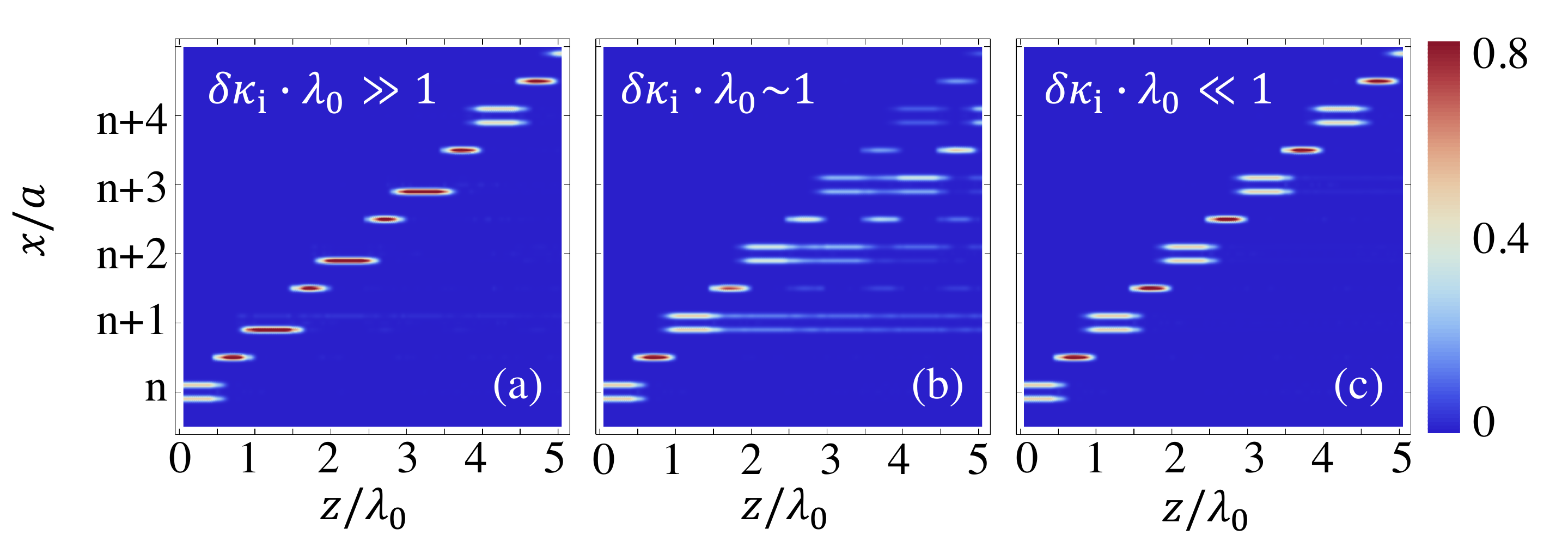}
%		\hspace{0.1cm}
%		\includegraphics[width=0.4\columnwidth]{Fig2-2.pdf}\\ 
%		\vspace{0.2cm}
%		\includegraphics[width=0.4\columnwidth]{Fig2-3.pdf} 
%		\hspace{0.1cm}
%		\includegraphics[width=0.4\columnwidth]{Fig2-4.pdf} 
		\caption{(a-c) Effects of disorder on Thouless pumping. Evolution of the field's intensity along the cycle ${\cal C}_1$ for different disorder strengths: (a) $\delta\kappa_i \lambda_0 \sim 50 $, (b) $\delta \kappa_i \lambda_0 \sim 1$, (c) $\delta\kappa_i \lambda_0\sim 0.1$. Other parameters as in Fig. \ref{noncom}} %Panels (a) and (b) show the non-commutativity of cycles  ${\cal C}_{1}$ and ${\cal C}_{2a}$. }
		\label{disorder}
	\end{center}
\end{figure}

So far we considered an idealized situation. Various effects may disturb Thouless pumping, such as non-linearity, non-adiabatic effects and disorder.
In Fig. \ref{disorder} we consider the effects of disorder, the latter may cause a  non-perfect periodicity of the lattice enabling transition between different Bloch states, it may distort pumping cycles or break the lattice Hamiltonian symmetries. To exemplify the effects of disorder in Fig. \ref{disorder} we show the evolution of the field intensity along the cycle ${\cal C}_1$  in the presence of impurities of different strengths, yielding the following correction to the Hamiltonian, $\delta H=\sum_i \delta \kappa_i (c^\dag_i c_i-d^\dag_i d_i)$. We see that as long as $\delta\kappa_i\lambda_0\ll1$, Fig. \ref{disorder}(c), pumping is not affected by the presence of impurities. In the opposite limit $\delta\kappa_i\lambda_0\gg1$, the modulation is adiabatic also compared to the splitting introduced by the impurity potential,  we thus see in Fig. \ref{disorder}(a) that during the pumping cycle the field intensity follows the impurity profile but the whole process does not cause dispersion or non-adiabatic transitions. Eventually, in the case $\delta\kappa_i\lambda_0\sim 1$ disorder couples the symmetric and antisymmetric modes, causing imperfections and leakage of Thouless pumping.
We note however that the non-abelian Lieb chain offer an advantage over its abelian version. Indeed,  since the degenerate bands of the naL lattice are flat, for any kind of modulation dispersive effects, that may hinder the observation of Thouless pumping, are vanishing.

\section{Non-adiabatic effects and photonic waveguide implementation}
\label{sect-photonics}
In this section we discuss in more detail the implementation of non-abelian photon pumping in photonic waveguide arrays.
We briefly explain how the tight-binding Hamiltonian emerge from Maxwell equations and we provide the relation between the tunnel couplings $J$  and the geometrical and optical system parameters. 
We then identify the requirements to realize  the adiabatic pumping regime  and we discuss possible deviations from the idealized adiabatic limit.

Electromagnetic wave propagation in a coupled-waveguide system should be investigated by solving the exact Maxwell equations in a medium with a spatially dependent refractive index.
However,  the analysis of such a system can as well, with high accuracy, be described 
by the coupled mode theory (CMT)\cite{Yariv,Snyder72,snyder,Marcuse}, a scheme that provides an easy understanding of the coupling mechanism and allows to reduce the problem to the solution of a set of coupled equations.
In this framework, the fields overlap between adjacent waveguides introduces a polarization term $\vec{P}$ which modifies the modes of the single $n$-th unperturbed waveguide, so that, for harmonic fields at frequency $\omega$ we have the following equations describing electromagnetic wave propagation in the  $n$-th waveguide
\begin{equation}
\begin{cases}
\nabla \times \vec{E}_n(r)=-i\omega \mu \vec{H}_n(r)\\
\nabla \times \vec{H}_n(r)=i\omega (\epsilon \vec{E}_n(r)+\vec{P}_n) 
\end{cases}   \label{eq:max}     
\end{equation}
where $\mu$ and $\varepsilon$ are  the permeability
and permittivity of the material, respectively.
A simple way\cite{Lara} to obtain the coupled mode equations is by defining the transversal forward propagating perturbed (p) and unperturbed (u) fields as:
\begin{equation}
\begin{cases}
\vec{E}_{np}(r)=\alpha_{n}(z)e^{i\omega t}\vec{e}_n(x,y)\\
\vec{E}_{nu}(r)=e^{i\beta_n z}e^{i\omega t}\vec{e}_n(x,y)
\end{cases} \label{eq:esp3}     
\end{equation}
and making use of the Lorentz reciprocity theorem.
In Eq. (\ref{eq:esp3}) $\vec{e}_n(x,y)$ is one of the orthonormal transverse modes of the n-th waveguide and $\alpha_{n}(z)$ the propagating field amplitude which is assumed to vary along the propagation direction (z-direction).

Making use of the reciprocity relations we have:
\bea
& & \int_{S} \nabla \cdot [\vec{E}_{np}(r) \times \vec{H}^*_{nu}(r)+\vec{E}^*_{nu}(r) \times \vec{H}_{np}(r)] dS=\nn\\
& &= -i \omega \int_{S} \vec{P}_n \cdot \vec{E}^*_{nu}(r) dS
\label{eq:esp4}     
\eea
where the integral is over the waveguide cross section.
Using the definition of the perturbed and unperturbed fields given in (\ref{eq:esp3}) and the orthonormality condition $\int_{S}[\vec{e}_n\times \vec{h}_m^*+\vec{e}_m^*\times \vec{h}_n]dS=2P_0 \delta_{nm}$ with $P_0$ denoting the normalized power, equation (\ref{eq:esp4}) can be recast as follows
\begin{equation}
(i\partial_z+\beta_n)\alpha_n(z)=\frac{\omega}{2P_o} \int_{S}\vec{P}_n\cdot \vec{e}_n^* dS
\label{eq:esp5}     
\end{equation}
In the polarization term two contributions can be identified coming from the local and coupling perturbations: $\vec{P}_n=\Delta\epsilon_n\alpha_n(z)\vec{e}_n+\sum_{j=1,j\ne n}^{N}\Delta\epsilon_n\alpha_j(z)\vec{e}_j$ where $\Delta\epsilon_n$ is the deviation from the unperturbed permittivity $\epsilon$ in the waveguide $n$ due to the presence of the others. This allows to rewrite equation (\ref{eq:esp5}) in the form:
\begin{equation}
(i\partial_z+\beta_n-\gamma_n)\alpha_n(z)=\sum_{j=1,j\ne n}^{N} \alpha_j(z)\frac{\omega}{2P_o}\int_{S}\Delta\epsilon_n\vec{e}_j\cdot \vec{e}_n^* dS
\label{eq:esp7}     
\end{equation}
where:
$\gamma_n=\frac{\omega}{2P_0} \int_{S}\Delta\epsilon_n\vec{e}_n\cdot \vec{e}_n^* dS$.

If only nearest neighbor interactions are considered, it allows to define the tunnel  couplings as:
\begin{equation}
J_{ij}=\frac{\omega}{2P_0}\int_{S}\Delta\epsilon_i\vec{e}_i^*\cdot \vec{e}_jdS
\label{eq:esp8}     
\end{equation}
For each of the four waveguides $i$ in a cell they arise because of the presence of a mode in the adjacent guides $j$ and it depends  only on the parameters in two coupled waveguides. Their strong dependence on the separation between the two waveguides, the dielectric constants of the waveguide core and on the shape and dimension of the waveguide cross section, gives the possibility to tune the couplings in a wide range.

In realistic experiments, based on femtosecond laser written glass structures, the waveguides sizes are typically of the order of 5-10 $\mu$m, while their separation is
of the order of 10 - 20 $\mu$m. Typical effective indexes and operating wavelenghts in vacuum are $n_{e}=1.5$ and $\lambda$=0.8 $\mu$m, giving rise to an effective optical potential for the single waveguide $V_o \simeq$ 1 meV and a field decay length of a few micrometers. Typical coupling values\cite{ke2016} in this case are of the order of $J \simeq  \times 10^{-3} {\rm \mu m}^{-1}$. Moreover, for the adiabatic condition to be fulfilled, we need to have a modulation wavelength $\lambda_0 J >>1$. A value $\lambda_0 =5 {\rm cm}$ yields  $\lambda_0 J \sim 50$ and it gives the possibility to perform 4 modulation cycles in waveguides 20 cm long; this is  a realistic length where loss effects can be neglected.
A detailed discussion of non-abelian holonomic effects starting from coupled mode theory in photonic waveguide lattices can be found in Ref.[\onlinecite{pinske2021}].
%Thus the adiabatic pump protocols, described in the main text, can be written into a waveguide array by periodically modulating or the refraction index or the separation between the waveguides, as a function of the propagation distance, z, which changes the effective coupling coefficients in our lattice with a four sites unit cell.
%
%\subsection{Non-adiabatic effects} 
 \begin{figure}[t!]
 	\begin{center}
 		\includegraphics[width=0.4\textwidth]{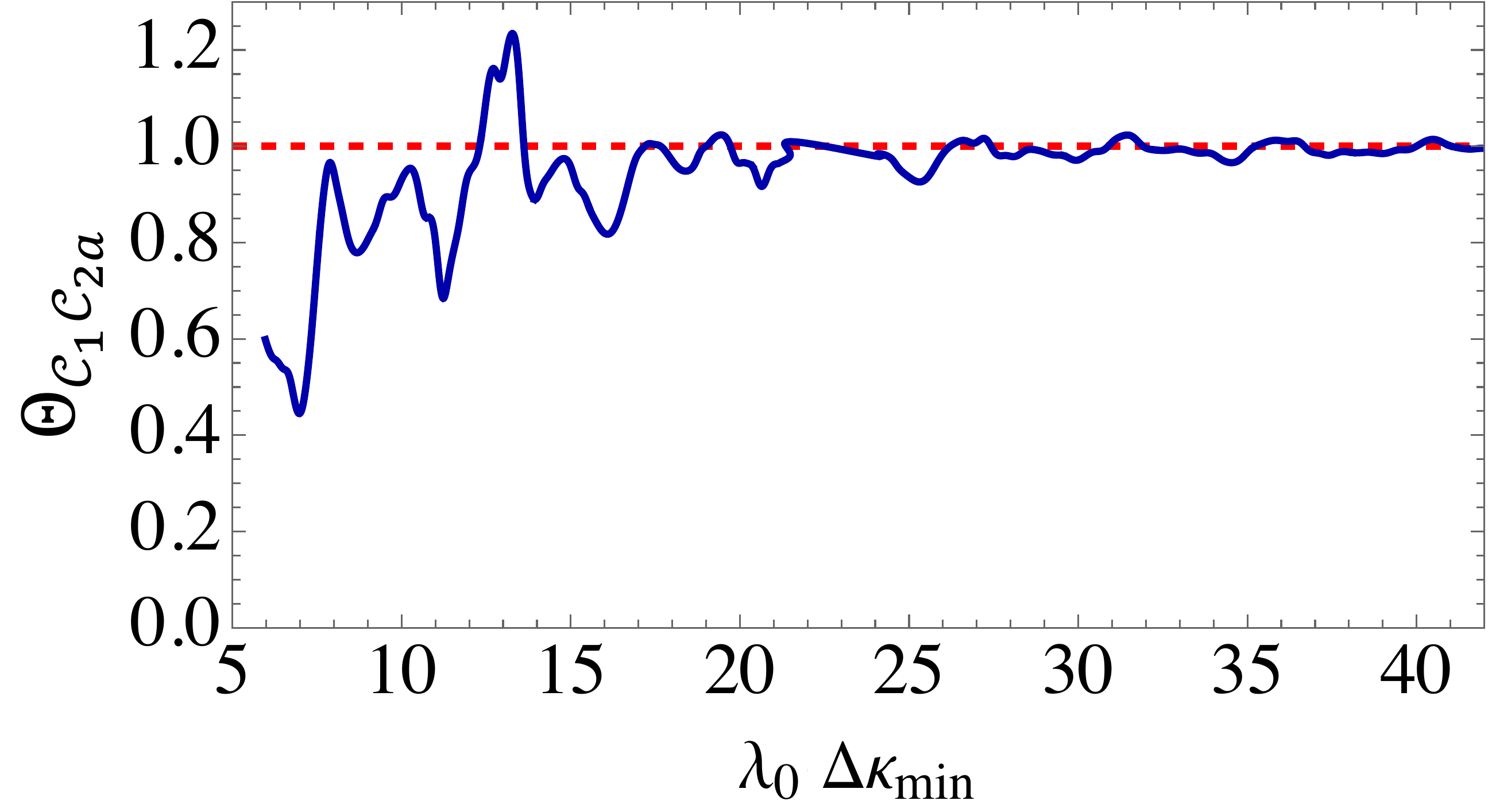}
 		\caption{Dependence  of the figure of merit $\Theta_{{\cal C}_1{\cal C}_{2a}}$ on the cycle wavelength, $\lambda_0$. The dashed lines $\Theta_{{\cal C}_1{\cal C}_{2a}}=1$ indicates the expected result in the perfectly adiabatic limit.}\label{nonad}
 	\end{center}
 \end{figure} 

Non-adiabatic effects may limit the possibility to isolate non-abelian holonomies. Differently from geometric contributions, non-adiabatic transitions will be however sensitive to the cycle duration and they could in principle be isolated performing a study of the figure of merit as a function of $\lambda_0$ as illustrated  in Fig.\ref{nonad}.
There we show how the figure of merit, $\Theta_{{\cal C}_1{\cal C}_{2a}}$ depends on the wavelength $\lambda_0$ the dashed line represent the asymptotic value in the adiabatic limit.
The wavelength is normalized to the minimum longitudinal momentum interval $\Delta \kappa_{\rm min}$ that is the minimum difference between the momenta of the Bloch eigenmodes.
Analogous methods can be used to isolate possible dispersive effects due to imperfections in the preparation of the initial state.

\section{conclusions}

Non-abelian gauge fields lie at the very heart of many modern physical theories. We need new experimental routes and observables to disclose the importance of the Wilczek and Zee holonomy. 
Recently, it was shown that they lead to non-abelian Bloch oscillations and they induce a topologically protected inter-band beating.\cite{diliberto2020}
We have shown that properly designed photonic lattices enable the control of the beam evolution by non-commutative fields, allowing to test experimentally a relation between  non-abelian holonomies and the displacement of the photon beam.
While standard Thouless pumping simply yields a displacement of the injected signal across the lattice, non-abelian Thouless pumping yields a displacement across the lattice and it generates a holonomic transformation among the different bands. Illustrating the peculiar geometrical meaning of Wannier center displacement in lattices with degenerate bands, the present work extends the known results of Resta and Vanderbilt \cite{resta2007} and hinting at the possibility to detect the signatures of Wilczek-Zee connection in solids furthermore, it points at the possibility to investigate the role of non-abelian holonomies also in quantum Hall experiments.
 
Non-abelian Thouless pumping can be generalized in several directions, including nonlinear effects, see for example Ref. [\onlinecite{smirnova2020}], or considering the propagation of non-classical light in non-abelian lattices. Both these possibilities are unexplored so far and open several new questions concerning - for example - the effect of the non-abelian holonomy on entanglement or the impact of nonlinearity in breaking the hidden symmetries. Non-abelian topological photonics may stimulate further developments and applications for classical and quantum information \cite{pinske2020} and tests of fundamental physics. \section{Acknowledgments}
The present research was supported by Sapienza Ateneo, QuantERA ERA-NET Co-fund (Grant No. 731473, QUOMPLEX), PRIN PELM (20177PSCKT), the H2020 PhoQus (Grant No. 820392) and by Regione Lazio (L. R. 13/08) through project SIMAP.
\\
Rosario Fazio acknowledges partial financial support from the Google Quantum Research Award. 
R. F. research has been conducted within the framework of the Trieste Institute for Theoretical Quantum Technologies (TQT). 

\appendix 

\section{Spectrum of the non-abelian Lieb chain}
\label{spectrum}
As discussed in Section \ref{sect-main},  a suitable basis to describe the propagation of electromagnetic waves through the modulated naL chain is the \textit{local Bloch eigenmodes basis} defined by  the following equations
\bea
&&
{\cal H}_k(z)|\psi_{\nu \a}(k,z)\ra=\kappa_\nu(z)|\psi_{\nu \a}(k,z)\ra\\
&& \la \psi_{\nu a}(k,z)|\psi_{\mu b}(k,z)\ra=\delta_{\mu\nu}\delta_{ab}
\eea
where  ${\cal H}_{k}$ is the Hamiltonian of the non-abelian Lieb chain at a given $k$
\be\label{Hk}
{\cal H}_{k}=
\left(
\begin{array}{cccc}
	0  & J_{b}(k)   & J_c & J_d   \\
	J_{b}^*(k)  &0   &0&0   \\
	J_c  &0   &0 &0 \\
	J_d &0&0&0
\end{array}
\right).
\ee
%where we also allow a different longitudinal propagation wave-vector between the waveguide $a$ and the waveguides $b,\, c$ and $d$, in the main text we set $\kappa=0$.
  The set $\{|\psi_{\nu a}(k,z)\ra\}$ yields the spectrum of the naL Hamiltonian,  and it features two dispersionless
 degenerate modes, $|\psi_{01}\ra$ and $|\psi_{02}\ra$, with longitudinal momentum, $\kappa_{0}=0$, given by Eq.\eqref{eq:basisNA2}   and two dispersive modes, $|\psi_{\pm}\ra$, defined as 
  \be
  \psi_{\pm}=\frac{1}{\sqrt{2}}\lf(|a_k\ra\pm \frac{J_b(k)|b_k\ra+J_c|c_k\ra+J_d|d_k\ra}{\Delta(k)}\rg)
 \ee
 with longitudinal momenta   $\kappa_{\pm}=\pm\Delta(k)$ where  $\Delta(k)=\sqrt{J^2_b(k)+J_c^2+J_d^2}$ and $J_b(k)=J_{b1}+J_{b2} e^{ik}$.
In the above equations  $|m_k\ra$  are the standard Bloch vectors,
 $|m_k\ra=\sum_i |m_i\ra e^{ikR_i}$ with $m=a,b,c,d$ and $ |m_i\ra$ indicating a state localized on the site $m$ of the cell $i$ located at $x=R_i$. %and we indicated with $N_{\nu a}$ the following normalization factors
% \be
\section{Non-commutative nature of the displacement generated in composite cycles}
As shown in Section \ref{sect-cycles}, starting from specific examples, the non-abelian nature of the evolution implies that, when we perform a sequence of two cycles, the  displacement depends on the ordering of the sequence.
To understand this fact, let us take two generic cycles, ${\cal C}_x$ and ${\cal C}_y$,  and the  two possible sequences of the two, namely, ${\cal C}_x{\cal C}_y$ and ${\cal C}_y{\cal C}_x$.
In the latter case, {\sl i.e.}  if we cover first the cycle ${\cal C}_x$ and then the cycle ${\cal C}_y$, the displacement matrix reads
%
% \bea\label{dcxcy}
%\lf[D_{{\cal C}_y{\cal C}_x}\rg]_{ab} &=& \lf[D_{{\cal C}_x}\rg]_{ab}+\\& & \!\!\!\!\!\!\!\!\!\! +\frac{1}{2\pi} \int dk\lf[W^{\dag}_{{\cal C}_x}\lf(\oint_{{\cal C}_y}W^{\dag}{\cal F}W\, dz\rg) W_{{\cal C}_x}\rg]_{ab}\nn\eea 
%%
%while in the opposite case it reads
%
 \be\label{dcycx}
\lf[D_{{\cal C}_x{\cal C}_y}\rg]_{ab} = \lf[D_{{\cal C}_y}\rg]_{ab}+\frac{1}{2\pi} \int dk\lf[W^{\dag}_{{\cal C}_y}{\hat D}_{{\cal C}_x}W_{{\cal C}_y}\rg]_{ab}\ee
while in the opposite case it reads
 \be\label{dcxcy}
\lf[D_{{\cal C}_y{\cal C}_x}\rg]_{ab} = \lf[D_{{\cal C}_x}\rg]_{ab}+\frac{1}{2\pi} \int dk\lf[W^{\dag}_{{\cal C}_x}{\hat D}_{{\cal C}_y}W_{{\cal C}_x}\rg]_{ab}\ee
where $W_{\cal C}$ and  ${ D}_{{\cal C}}$ indicate respectively the holonomy and the displacement matrix of the cycle ${\cal C}$ while  ${\hat D}_{{\cal C}}$  denotes the $k$-dependent displacement matrix  of the cycle ${\cal C}$, {\sl i.e.}  ${\hat D}_{{\cal C}}=\oint_{{\cal C}}W^{\dag}{\cal F}W\, dz $.
%while in the opposite case it reads
%%
%\bea\label{dc3c4}
%\lf[D_{{\cal C}_x{\cal C}_y}\rg]_{ab} &=& \lf[D_{{\cal C}_y}\rg]_{ab}+\\& & \!\!\!\!\!\!\!\!\!\! +\frac{1}{2\pi} \int dk\lf[W^{\dag}_{{\cal C}_y}\lf(\oint_{{\cal C}_x}W^{\dag}{\cal F}W\, dz\rg) W_{{\cal C}_y}\rg]_{ab}\nn\eea 
%where we denoted with $ (D_{\ldots})_{\a\b}$ and
 From the above equations we see that two cycles performed in different orders generate in general different displacements, due to the non-commutativity of  $W_{\cal C}$ and ${\hat D}_{{\cal C}}$.
%The same  result is also hinted at by the fact that performing a sequence of cycles in different orders yields a different holonomy and thus a different final state.
% in the basis defined by equations (\ref{scheme}). \\

This fact is true in particular for the cycles  ${\cal C}_{1}$ and  ${\cal C}_{2a}$ or ${\cal C}_{2b}$ presented in Section \ref{sect-cycles}.
We notice that, independently on the initial state the displacement accumulated along the cycles ${\cal C}_{2a}$ or ${\cal C}_{2b}$  is zero, since along
these cycles  the coupling $J_{b1}$  vanishes, {\sl i.e.} we have 
\be \label{condition} D_{{\cal C}_{2a}}=D_{{\cal C}_{2b}}=0 \ee%
%This physically transparent result can also be verified with an explicit calculation.
 Substituting the previous relations in Eqs.(\ref{dcycx}-\ref{dcxcy}) we obtain the following result for the photon displacement for the composite cycles ${\cal C}_{2i}{\cal C}_1$ and $ {\cal C}_1{\cal C}_{2i}$ 
\be\label{DTOT1} D_{{\cal C}_{2i}{\cal C}_1}=D_{{\cal C}_1}\ee
and 
\be\label{DTOT2} D_{{\cal C}_1{\cal C}_{2i}}=\frac{1}{2\pi} \int dk\lf[W^{\dag}_{{\cal C}_{2i}}{\hat D}_{{\cal C}_1}W_{{\cal C}_{2i}}\rg]\ee
with $i=a\,{\rm or}\,b$.

\section{Holonomies, field-strength and  displacement of the cycles of  ${\cal C}_1$,  ${\cal C}_{2a}$ and ${\cal C}_{2b}$}
\label{app-cycles}
In this Appendix we relate the holonomies and displacements generated in the cycles ${\cal C}_1$,  ${\cal C}_{2a}$ and ${\cal C}_{2b}$ to the geometrical structure of these cycles. 
Most numerical results presented in Section \ref{sect-cycles} will be explained analytically. 

%and the crucial features of the cycles that should be preserved to see non-commutative effects are identified.
%The geo figure of merit introduced in 
%
%The cycles ${\cal C}_1$,   ${\cal C}_{2a}$ and ${\cal C}_{2b}$ presented in Section \ref{sect-cycles} are sufficiently simple that the corresponding holonomies and displacements can be calculated analytically.
%%Since most results are independent on wether we consider ${\cal C}_{2a}$ or ${\cal C}_{2b}$ we indicate both these cycles simply as ${\cal C}_2$.
We start by estimating the holonomies, $W_{{\cal C}_1}$, $W_{{\cal C}_{2i}}$ with $i=a,\,b$.
The general expression of the  connection matrix, $\Gamma^z_0$, assuming a generic dependence of all tunnel coupling $J_i$ on $z$,  is 
\bea \label{explicitgammaz}\Gamma^z_0&=&\frac{
\lf(J_{b2}\partial_z{J}_{b1}-J_{b1}\partial_z{J}_{b2}\rg)\sin k}{\Delta^2(k)}\left(%
\begin{array}{cc}
  0  & 0 \\
  0 & 1 \\
\end{array}%
\right)+\nn\\& &+ \frac{
\lf(J_c\partial_zJ_d-J_d\partial_zJ_c\rg)|J_b|}{\delta^2\Delta(k)}\left(%
\begin{array}{cc}
  0  & i e^{i\phi_k}\\
  -i e^{-i\phi_k}& 0 \\
\end{array}%
\right),\nn\\ \eea
where  we set $\phi_k=\arg(J_{b1}+J_{b2} e^{ik})$
The connection along $k$ is instead given by
\bea  \label{explicitgammak}\Gamma_0^k&=& -\frac{
J_{b2}\lf(J_{b2}+J_{b1} \cos k\rg)}{\Delta^2(k)}\left(%
\begin{array}{cc}
  0  & 0 \\
  0 & 1\\
\end{array}%
\right).\eea
From the above equations we can easily evaluate  the holonomies of the cycles ${\cal C}_1$ and ${\cal C}_{2i}$ with $i=a,b$, we obtain
\bea\label{wc1}
W_{{\cal C}_1}&=&\exp\lf[i\Delta_{{\cal C}_1}\frac{1}{2}\lf(\sigma_0-\sigma_z\rg)\rg]\\[0.1cm]
W_{{\cal C}_{2i}}&=&\exp\lf[i\Delta_{{\cal C}_{2i}}\lf( \sin(k)\sigma_x+\cos(k)\sigma_y\rg)\rg]\label{wc2}
\eea
with $\sigma_0$ denoting the $2\times 2$ identity matrix and $\sigma_x,\,\sigma_y$ and $\sigma_z$ the three standard Pauli matrices. 
To derive equations (\ref{wc1}-\ref{wc2}), we used the fact that, along the paths ${\cal C}_{2i}$, $J_{b2}$ is constant and $J_{b1}=0$  and we indicated with $\Delta_{{\cal C}_1}$ and $\Delta_{{\cal C}_{2i}}$ the following integrals
\bea
& & \Delta_{{\cal C}_1}= \oint_{{\cal C}_1} \frac{
\lf(J_{b1}\partial_z{J}_{b2}-J_{b2}\partial_z{J}_{b1}\rg)\sin k}{J_d^2+J_c^2+|J_b|^2},\label{deltac1}\\[0.1cm] 
& & \Delta_{{\cal C}_{2i}}= \oint_{{\cal C}_{2i}} \frac{
\lf(J_c\partial_zJ_d-J_d\partial_zJ_c\rg)J_{b2}}{(J_c^2+J_d^2)(J_c^2+J_d^2+J_{b2}^2)^{1/2}}.\label{deltac2}
\eea
Both integrals can be easily calculated and they have a simple geometrical meaning.
 \begin{figure}[t!]
 \includegraphics[width=0.5\textwidth]{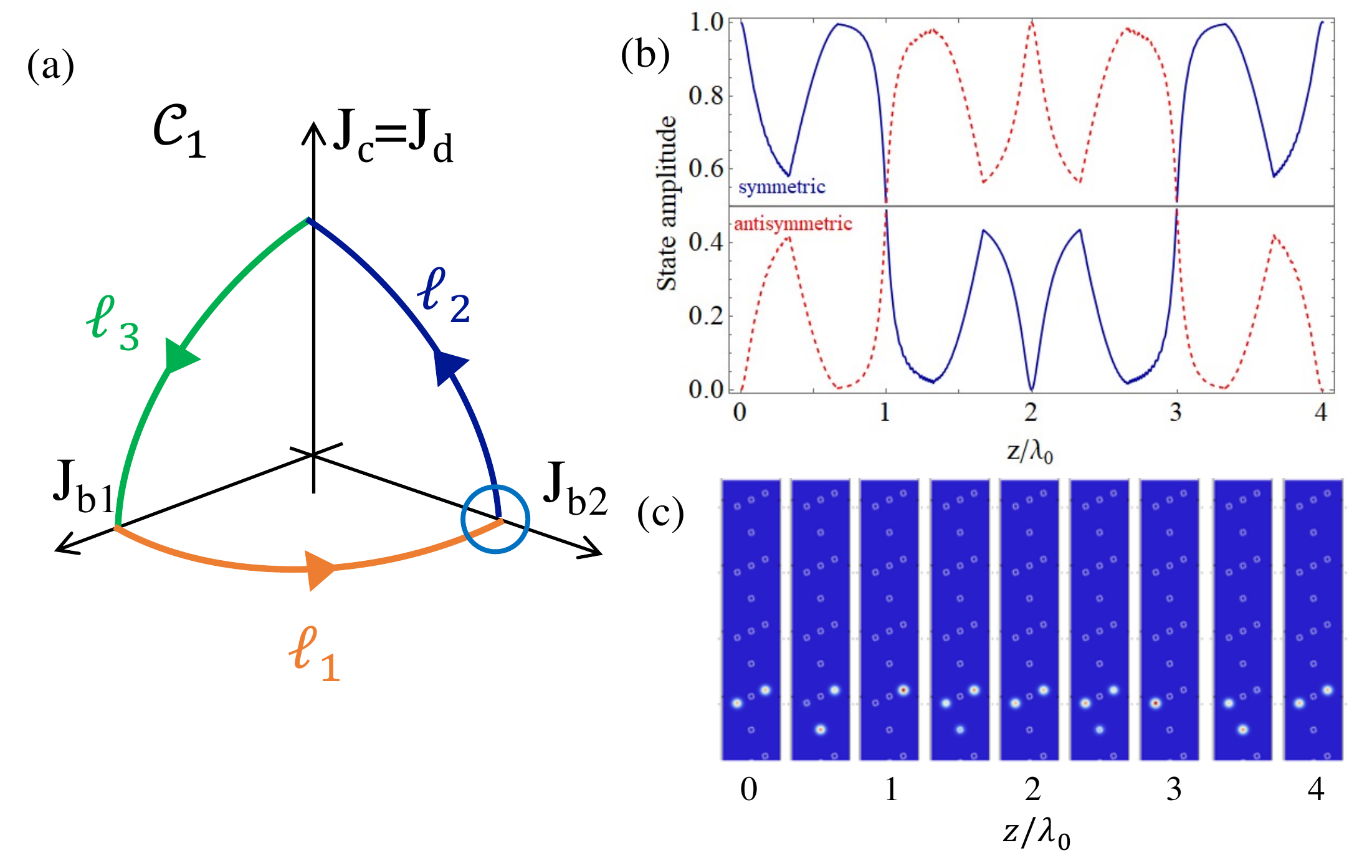}
 		\caption{(a) Structure of the cycle ${\cal C}_1$. (b) Evolution of the occupation of the two Bloch eigenmodes along the cycle ${\cal C}_{2a}$, the solid and dotted lines correspond respectively to the states $|\psi_{02}\ra$ and $|\psi_{01}\ra$. (c) Density plots of the field intensity across the lattice at different $z$ along the cycle ${\cal C}_{2a}$. }\label{structureandev}
 \end{figure} 
Specifically, for $\Delta_{{\cal C}_1}$ we have:
\be
\Delta_{{\cal C}_1}=\int_{\ell_1} \partial_z \arg\phi_k=k
\ee
where $\ell_1$ denotes the portion of the curve ${{\cal C}_1}$ laying on the $J_{b1},J_{b2}$ plane, see Fig. \ref{structureandev}(a).

Along the cycle ${\cal C}_1$ the occupation of the two eigenstates does not change however, the structure of the state $\psi_{02}$ is deformed in such a way that its Wannier center  is displaced by one unit cell per cycle as one can clearly see in Figs.\ref{noncom}(a) and \ref{displ}(a). 

At variance with $\Delta_{{\cal C}_1}$, the integral $\Delta_{{\cal C}_{2i}}$ depends on the precise shape of the cycle. For the two examples shown in Fig.2(b) of the main text the calculation is straightforward; specifically, we get $\Delta_{{\cal C}_{2i}}=\pi/2-\alpha_{i}$ with $\alpha_a=\pi/4$ and $\alpha_b=\pi/3$ where the $\pi/2$ contribution comes in both cases from the fact that we assume an infinitesimal value of $J_c=J_d$ at the starting point to avoid ambiguities in the definition of the states. We notice that $\Delta_{{\cal C}_{2i}}$ has a simple geometric interpretation, indeed, denoting with $\bar{\cal C}_{2i}$ the projection of the path ${\cal C}_{2i}$ on the unitary sphere in the space $J_c$,$J_d$, $J_{b2}$, one can show that $\Delta_{{\cal C}_{2i}}$ yields the solid angle subtended by the path $\bar{\cal C}_{2i}$ at the origin.
\\
To see this we introduce the cartesian coordinates $\{x,y,z\}=\{J_c,J_d,J_{b2}\}$ and we rewrite $\Delta_{{\cal C}_{2i}}$ as:
\be \Delta_{{\cal C}_{2i}}=\int_{{\cal C}_{2i}}
\frac{z\lf(ydx-x\,dy\rg)}{(x^2+y^2)\sqrt{x^2+y^2+z^2}}.\label{contour-int}\ee
Noting that
$$\nabla\times\lf[\frac{z\lf(-y,x,0\rg)}{(x^2+y^2)\sqrt{x^2+y^2+z^2}}\rg]=\frac{(x,y,z)}{\lf(x^2+y^2+z^2\rg)^\frac{3}{2}}$$  and applying Stokes theorem we can then rewrite the contour integral in Eq.\eqref{contour-int} as a surface integral as follows
\be \Delta_{{\cal C}_{2i}}=\int_{\Sigma_i}\frac{\vec r}{r^{3}}\cdot
d\vec\sigma\ee
where we denoted with $\vec r$ the vector $\lf(x,y,z\rg)$ and with $\Sigma_i$ a surface such that $\partial\Sigma_i={\cal C}_{2i}$. Eventually, indicating as $\bar\Sigma_i$ the portion of the unitary sphere,$S^2$, enclosed by $\bar {\cal C}_{2i}$, with an appropriate choice of $\Sigma_i$  we can easily obtain the desired result, {\sl i.e.}
\be \Delta_{{\cal C}_{2i}}=\int_{\bar \Sigma_i}\frac{\vec r}{r^{3}}\cdot
d\vec\sigma=\int_{\bar \Sigma_i}d\o\ee
where we have used the fact that, on $S^2$, $d\vec\sigma=d\o\hat r$ with $d\o$ the differential of solid angle.
In Figure \ref{structureandev}(b) we show the evolution of the population of the two eigenstates along the cycle ${\cal C}_{2a}$ , as one can see this cycle yields a rotation by $3\pi/4$.

We now focus on the displacement matrices.
To this end, we employ Eq.\eqref{eq:displ-matr}, that relates the displacement matrix to the field strength.  We thus start from  the following general expression of  the field strength matrix, ${\cal F}$, 
\begin{widetext}
\be \label{expliciteffe}{\cal F}_{zk}=\frac{
\lf(J_c\partial_zJ_d-J_d\partial_zJ_c\rg)J_{b2}}{\Delta^3(k)}\left(%
\begin{array}{cc}
  0  &e^{ik}\\
  e^{-i k}& 0 \\
\end{array}%
\right)+\frac{
J_{b2} \lf[(J_{b2}+J_{b1} \cos(k))\partial_z\delta^2-\delta^2\partial_z(J_{b2}+J_{b1} \cos(k))\rg]}{\Delta^4(k)}\left(%
\begin{array}{cc}
  0  & 0 \\
  0 & 1 \\
\end{array}%
\right). \ee
\end{widetext}
where we  assumed that all tunnel couplings are generic function of $z$.

Analogously to Eqs.(\ref{explicitgammaz}-\ref{explicitgammak}), Eq.\eqref{expliciteffe} substantially simplifies if we consider the geometrical structure of the cycles.

Along the cycle ${\cal C}_1$ we have $J_c=J_d$, the first term in Eq.\eqref{expliciteffe} is thus vanishing  and the field strength ${\cal F}$ can be cast as 
\be {\cal F}_{zk}\xrightarrow[\text{along ${\cal C}_1$}]{\text{{ }}}\frac{2J_{b2} J_c(J_{b2}\partial_z J_c-J_c\partial_z J_{b2})}{J_{b2}^2+J_c^2} \lf(\begin{array}{cc}
  0  & 0 \\
  0 & 1 \\
\end{array}%
\right) \ee
This term is non-vanishing only along the portion $\ell_2$ of the cycle ${\cal C}_1$,  it does not depend on $k$ and it commutes with the holonomy $W_{{\cal C}_1}$. Furthermore,  as one can easily check, integrated over $z$ and $k$ yields $2\pi$. It thus leads to the following expression for the displacement matrix of the cycle ${\cal C}_1$
\be D_{{\cal C}_1}=\left(%
\begin{array}{cc}
  0 & 0 \\
  0 & 1 \\
\end{array}%
\right).\ee
As explained  above, the displacement generated along the cycles ${\cal C}_{2i}$ is zero  nevertheless, by looking at equation \eqref{expliciteffe}
we notice that along these cycles we have a finite field strength. As one can explicitly verify, the zero displacement results is indeed recovered
after integrating over $k$ and $z$.
In these regards we note that for $J_{b2}=0$ the field strength is identically zero, opposite to what happens for $J_{b1}=0$, 
Since, due to the lattice periodicity I can always exchange the roles of  $J_{b1}$ and $J_{b2}$ and I expect that the symmetry between the two cases is restored upon averaging over $k$,
this is indeed the case.
 
A possible route to demonstrate the non-commutative nature of non-abelian Thouless pumping  is to design a  pumping cycle aimed at detecting the contribution to the displacement  of the commutator term 
${\cal F}^{\rm com}_{zk}=i\lf[\Gamma_z,\Gamma_k\rg]$,  peculiar of non-abelian gauge theories. We remark that, in spite of the apparent complexity, in Eq. \eqref{explicitgammaz} we have a cancellation between terms coming from the abelian contribution to the field strength, $\partial_k \Gamma^z_\nu-\partial_z \Gamma^k_\nu$ and  from the non-abelian commutator,  
${\cal F}^{\rm com}_{zk}$. It is interesting to look at the explicit expression of the latter

\bea {\cal F}^{\rm com}_{zk}&=&\frac{
\lf(J_c\partial_zJ_d-J_d\partial_zJ_c\rg)|J_b|J_{b2}(J_{b2}+J_{b1}\cos(k))}{\delta^2\Delta^3(k)} \cdot \nn\\ & & \cdot \left(%
\begin{array}{cc}
  0  & i e^{i\phi_k}\\
  -i e^{-i\phi_k}& 0 \\
\end{array}%
\right),\label{fcom}\eea
by looking at the above equation we see that  if $J_{b1}$ or   $J_{b2}$ are identically zero along the cycle the contribution of this terms vanishes upon averaging over $k$. To see the contribution of this term in pumping we thus need to design a cycle having all the coupling different from zero at some point.

\section{Non-commuting cycles sequences}
Using the above results, the holonomies for the composite cycles  ${\cal C}_1{\cal C}_{2}$ and ${\cal C}_{2}{\cal C}_1$ can be explicitly calculated. They are given by
\be\label{W1W2}
W_{{\cal C}_1{\cal C}_{2}}=\lf(\begin{array}{cc}
\cos(\Delta_{{\cal C}_{2}}) & e^{ik}\sin(\Delta_{{\cal C}_{2}})  \\
-\sin(\Delta_{{\cal C}_{2}})  & e^{ik} \cos(\Delta_{{\cal C}_{2}}) \\
\end{array}\rg)%
\ee
and 
\be\label{W2W1}
 W_{{\cal C}_{2}{\cal C}_1}=\lf(\begin{array}{cc}
\cos(\Delta_{{\cal C}_{2}}) & e^{2ik}\sin(\Delta_{{\cal C}_{2}})  \\
-e^{-ik}\sin(\Delta_{{\cal C}_{2}})  & e^{ik} \cos(\Delta_{{\cal C}_{2}}) \\
\end{array}\rg)%
\ee
The corresponding displacement matrices can be instead expressed as follows:
\bea D_{{\cal C}_{2}{\cal C}_1}&=&D_{{\cal C}_1}=\left(%
\begin{array}{cc}
  0 & 0 \\
  0 & 1 \\
\end{array}%
\right)\label{D1} \\
D_{{\cal C}_1{\cal C}_{2}}&= & \left(%
\begin{array}{cc}
  \sin^2\D_{{\cal C}_{2}} & 0 \\
 0 & \cos^2{\D_{{\cal C}_{2}}} \\
\end{array}\right)\label{D2}
\eea
 Equations (\ref{W1W2}-\ref{W2W1}) ashow that the holonomy transformations  for the cycles ${\cal C}_1$ and  ${\cal C}_{2}$  do not commute, {\sl i.e.}
 starting from a given initial state, the final state and  the displacement obtained depend on wether we perform first the cycle ${\cal C}_1$ and then the cycle ${\cal C}_{2}$ or {\sl viceversa}.
  In the table below we explicitly calculate the final state and displacement generated starting from different localized initial states. 
Specifically, we consider the case where the initial states are simply $|c_n\ra$ and $|d_n\ra$, and get:
\begin{widetext}
\be
\begin{tabular}{|c|c|c|c|}
\hline
Cycles& Initial state & Final state & Displacement\\
    \hline
    \multirow{3}{*}{${\cal C}_{2}{\cal C}_1$}& & & \\[-0.3cm]
    &$|c_n\ra$&$\eta_+ \lf(|c_n\ra+|d_{n+1}\ra\rg)+\eta_-(|c_{n+1}\ra-|d_n\ra) $&$\frac{1}{2}$\\[0.2cm]
    &$|d_n\ra$&$\eta_+ (|d_{n+1}\ra-|c_n\ra)+\eta_-(|d_n\ra+|c_{n+1}\ra) $&$\hf$\\[0.2cm]    \hline
     \multirow{3}{*}{${\cal C}_1{\cal C}_{2}$}& & & \\[-0.3cm]
   &$|c_n\ra$&$\eta_- (|c_n\ra-|d_n\ra)+\eta_+(|c_{n+1}\ra+|d_{n+1}\ra) $&$\frac{1+\sin(2\D_{\cc})}{2}$\\[0.2cm]
    &$|d_n\ra$&$\eta_+ (|d_n\ra-|c_n\ra)+\eta_-(|d_{n+1}\ra+|c_{n+1}\ra) $&$\frac{1-\sin(2\D_{\cc})}{2}$\\[0.2cm]    \hline
  \hline
\end{tabular}
\ee
\end{widetext}
 with $2\eta_{\pm}=\cos(\D_{\cc})\pm\sin(\D_{\cc})$.
  %

%\subsection{Non-adiabatic effects} 
% \begin{figure}[h!]
% 	\begin{center}
% 		\includegraphics[width=0.4\textwidth]{Figure-NON-AD}
% 		\caption{Dependence  of the figure of merit $\Theta_{{\cal C}_1{\cal C}_{2a}}$ on the cycle wavelength, $\lambda_0$. The dashed lines $\Theta_{{\cal C}_1{\cal C}_{2a}}=1$ indicates the expected result in the perfectly adiabatic limit.}\label{nonad}
% 	\end{center}
% \end{figure} 
%
%Non-adiabatic effects may limit the possibility to isolate non-abelian holonomies. Differently from geometric contributions, non-adiabatic transitions will be however sensitive to the cycle duration and they could in principle be isolated performing a study of the figure of merit as a function of $\lambda_0$ as illustrated  in Fig.\ref{nonad}.
%There we show how the figure of merit, $\Theta_{{\cal C}_1{\cal C}_{2a}}$ depends on the wavelength $\lambda_0$ the dashed line represent the asymptotic value in the adiabatic limit.
%The wavelength is normalized to the minimum longitudinal momentum interval $\Delta \kappa_{\rm min}$ that is the minimum difference between the momenta of the Bloch eigenmodes.
%


\begin{thebibliography}{10}

\bibitem{bott1965}
R. Bott and S.~S. Chern, Acta Mathematica {\bf 114},  71  (1965).

\bibitem{simon1983}
B. Simon, Phys. Rev. Lett. {\bf 51},  2167  (1983).

\bibitem{berry1984}
M.~V. Berry, Proceedings of the Royal Society of London. A. Mathematical and
  Physical Sciences {\bf 392},  45  (1984).

\bibitem{wilczek1984}
F. Wilczek and A. Zee, Phys. Rev. Lett. {\bf 52},  2111  (1984).

\bibitem{wilczek1989}
F. Wilczek and A. Shapere, Geometric Phases in Physics, World Scientific,
  Singapore, 1989.

\bibitem{resta2007}
R. Resta and D. Vanderbilt,  in  Physics of Ferroelectrics, edited by
  K.~M. Rabe, C.~H. Ahn, and J.-M. Triscone (Springer-Verlag GmbH, Berlin,
  2007), Chap.~Theory of Polarization: A Modern Approach.

%\bibitem{hasan2010}
%M.~Z. Hasan and C.~L. Kane, Rev. Mod. Phys. {\bf 82},  3045  (2010).
%
%%\bibitem{ozawa2019a}
%T. Ozawa and H.~M. Price, Nature Reviews Physics {\bf 1},  349  (2019).



\bibitem{thouless1982}
D.~J. Thouless, M. Kohmoto, M.~P. Nightingale, and M. den Nijs, Phys. Rev.
  Lett. {\bf 49},  405  (1982).

\bibitem{thouless1983}
D.~J. Thouless, Phys. Rev. B {\bf 27},  6083  (1983).

\bibitem{aunola2003}
M. Aunola and J.~J. Toppari, Phys. Rev. B {\bf 68},  020502  (2003).

\bibitem{mottonen2006}
M. M\"ott\"onen, J.~P. Pekola, J.~J. Vartiainen, V. Brosco, and F.~W.~J. Hekking,
  Phys. Rev. B {\bf 73},  214523  (2006).

\bibitem{xiao2010}
D. Xiao, M.-C. Chang, and Q. Niu, Rev. Mod. Phys. {\bf 82},  1959  (2010).

\bibitem{zilberberg2018}
O. Zilberberg, S. Huang, J. Guglielmon, M. Wang, K.~P. Chen, Y.~E. Kraus, and
  M.~C. Rechtsman, Nature {\bf 553},  59  (2018).

\bibitem{PhysRevB.93.245113}
H.~M. Price, O. Zilberberg, T. Ozawa, I. Carusotto, and N. Goldman, Phys. Rev.
  B {\bf 93},  245113  (2016).

\bibitem{Aidelsburger2014}
M. Aidelsburger, M. Lohse, C. Schweizer, M. Atala, J.~T. Barreiro, S.
  Nascimb{\`{e}}ne, N.~R. Cooper, I. Bloch, and N. Goldman, Nature Physics {\bf
  11},  162  (2014).

\bibitem{kraus2012}
Y.~E. Kraus, Y. Lahini, Z. Ringel, M. Verbin, and O. Zilberberg, Phys. Rev.
  Lett. {\bf 109},  106402  (2012).

\bibitem{ke2016}
Y. Ke, X. Qin, F. Mei, H. Zhong, Y.~S. Kivshar, and C. Lee, Laser {\&}
  Photonics Reviews {\bf 10},  1064  (2016).

\bibitem{wimmer2017}
M. Wimmer, H.~M. Price, I. Carusotto, and U. Peschel, Nat. Phys. {\bf 13},  545
   (2017).

\bibitem{fedorova2019}
Z. Fedorova, C. Jörg, C. Dauer, F. Letscher, M. Fleischhauer, S. Eggert, S.
  Linden, and G. von Freymann, Light: Science {\&} Applications {\bf 8},  63
  (2019).

\bibitem{wang2019}
Y. Wang, Y.-H. Lu, F. Mei, J. Gao, Z.-M. Li, H. Tang, S.-L. Zhu, S. Jia, and
  X.-M. Jin, Phys. Rev. Lett. {\bf 122},  193903  (2019).
  
\bibitem{lohse2015}
 M. Lohse, C. Schweizer, O. Zilberberg, M. Aidelsburger, and I. Bloch, Nat.
 Phys. {\bf 12},  350  (2015).
  
\bibitem{nakajima2016}
 S. Nakajima, T. Tomita, S. Taie, T. Ichinose, H. Ozawa, L. Wang, M. Troyer, and
 Y. Takahashi, Nat. Phys. {\bf 12},  296  (2016).
  
  \bibitem{kremer2020}
M. Kremer, I. Petrides, E. Meyer, M. Heinrich, O. Zilberberg, and A. Szameit,
  Nature Communications {\bf 11},  907  (2020).

\bibitem{lieb1989}
E.~H. Lieb, Phys. Rev. Lett. {\bf 62},  1201  (1989).

\bibitem{vicencio2015}
R.~A. Vicencio, C. Cantillano, L. Morales-Inostroza, B. Real, C.
  Mej{\'{\i}}a-Cort{\'{e}}s, S. Weimann, A. Szameit, and M.~I. Molina, Phys.
  Rev. Lett. {\bf 114},  245503  (2015).

\bibitem{mukherjee2015}
S. Mukherjee, A. Spracklen, D. Choudhury, N. Goldman, P. \"Ohberg, E. Andersson,
  and R.~R. Thomson, Phys. Rev. Lett. {\bf 114},  245504  (2015).

\bibitem{duan2001}
L.-M. Duan, Science {\bf 292},  1695  (2001).

\bibitem{taddia2017}
L. Taddia, E. Cornfeld, D. Rossini, L. Mazza, E. Sela, and R. Fazio, Phys. Rev.
  Lett. {\bf 118},  230402  (2017).

\bibitem{faoro2003}
L. Faoro, J. Siewert, and R. Fazio, Phys. Rev. Lett. {\bf 90},  028301  (2003).

\bibitem{brosco2008}
V. Brosco, R. Fazio, F.~W.~J. Hekking, and A. Joye, Phys. Rev. Lett. {\bf 100},
   027002  (2008).

\bibitem{pachos2000}
J. Pachos and S. Chountasis, Phys. Rev. A {\bf 62},  052318  (2000).

\bibitem{iadecola2016}
T. Iadecola, T. Schuster, and C. Chamon, Phys. Rev. Lett. {\bf 117},  073901
  (2016).

\bibitem{chen2019}
Y. Chen, R.-Y. Zhang, Z. Xiong, Z.~H. Hang, J. Li, J.~Q. Shen, and C.~T. Chan,
  Nat. Commun. {\bf 10},  1  (2019).


\bibitem{zwanziger1990}
J.~W. Zwanziger, M. Koenig, and A. Pines, Phys. Rev. A {\bf 42},  3107  (1990).

\bibitem{martinis2020}
J.~M. Martinis, M.~H. Devoret, and J. Clarke, Nat. Phys. {\bf 16},  234
  (2020).

\bibitem{abdumalikov2013}
A.~A. Abdumalikov, J.~M. Fink, K. Juliusson, M. Pechal, S. Berger, A. Wallraff,
  and S. Filipp, Nature {\bf 496},  482  (2013).
  

\bibitem{ozawa2019}
T. Ozawa, H.~M. Price, A. Amo, N. Goldman, M. Hafezi, L. Lu, M.~C. Rechtsman,
  D. Schuster, J. Simon, O. Zilberberg, and I. Carusotto, Rev. Mod. Phys. {\bf
  91},  015006  (2019).

\bibitem{christodoulides2003}
D.~N. Christodoulides, F. Lederer, and Y. Silberberg, Nature {\bf 424},  817
  (2003).

\bibitem{snyder}A. W. Snyder and J. Love, \textit{Optical Waveguide theory}, Springer, Berlin, 1983. 

%\bibitem{sup}
%See Supplementary Material for more details.
%
  \bibitem{PhysRevResearch.1.033117}
M. Kremer, L. Teuber, A. Szameit, and S. Scheel, Phys. Rev. Research {\bf 1},
  033117  (2019).


\bibitem{casteels2016}
W. Casteels, R. Rota, F. Storme, and C. Ciuti, Phys. Rev. A {\bf 93},  043833
  (2016).

\bibitem{biondi2018}
M. Biondi, G. Blatter, and S. Schmidt, Phys. Rev. B {\bf 98},  104204  (2018).
%
%\bibitem{rice1982}
%M.~J. Rice and E.~J. Mele, Phys. Rev. Lett. {\bf 49},  1455  (1982).
%
%\bibitem{su1979}
%W.~P. Su, J.~R. Schrieffer, and A.~J. Heeger, Phys. Rev. Lett. {\bf 42},  1698
%  (1979).

\bibitem{bra-ket}
Coherently with the quantum interpretation of the classical Maxwell equations
  we use a bra-ket notation.

%\bibitem{adiabatic}
%Non-adiabatic effects may clearly hinder the possibility to isolate non-abelian
%  effects. Further discussion can be found in the Supplementary.

\bibitem{zak1989}
J. Zak, Physical Review Letters {\bf 62},  2747  (1989).

\bibitem{fishbane1981}
P.~M. Fishbane, S. Gasiorowicz, and P. Kaus, Phys. Rev. D {\bf 24},  2324
  (1981).

\bibitem{manton2004}
N. Manton and P. Sutcliffe, Topological Solitons (Cambridge University
  Press, Cambridge, UK, 2004).
  
 \bibitem{smirnova2020} 
 D. Smirnova,   D. Leykam, Y. Chong, and   Y. Kivshar,
 Appl. Phys. Rev. {\bf 7}, 021306 (2020).
  

  
\bibitem{Qianqian2020}
 Q. Chen, J. Cai and S. Zhang, Phys. Rev. A {\bf 101}, 043614 (2020)

\bibitem{Cerjan2020} 
 A. Cerjan, M. Wang, S. Huang, et al. Light. Sci. Appl. {\bf 9}, 178 (2020).
 
 \bibitem{Yariv}  
	A. Yariv, IEEE
	J. Quantum Electron., {\bf 9}, 919 (1973).
\bibitem{Snyder72}
	A. Snyder, J. Opt.
	SOC. Amer., {\bf  62}, 1267 (1972).
\bibitem{Marcuse}
	D. Marcuse,  Bell. Sys. Tech. J. {\bf 52}, 817 (1973)
\bibitem{pinske2021} J. Pinske and S. Scheel arXiv:2105.04859 (2021).
%\bibitem{Snyder83}  
%	A. W. Snyder and J.D. Love, "Optical Waveguide Theory", Chapman and Hall, 1983 
\bibitem{Lara} 
	B. M. Rodriguez-Lara, Francisco Soto-Eguibar, Demetrios N. Christodoulides, Phys. Scr. {\bf 90}, 068014 (2015)
\bibitem{diliberto2020} M. Di Liberto, N. Goldman and G. Palumbo, Nature Communications {\bf 11}, 5942 (2020)
\bibitem{pinske2020}
J. Pinske, L. Teuber, and S. Scheel, Phys. Rev. A {\bf 101}, 062314 (2020)




%\bibitem{Ke}	
%	Ke, Y., Qin, X., Mei, F., Zhong, H., Kivshar, Y.S. and Lee, C., Laser Photonics Reviews, {\bf 10}, 995 (2016)
%	
%  \bibitem{fishbane1981}
%	P.~M. Fishbane, S. Gasiorowicz, and P. Kaus, Phys. Rev. D {\bf 24},  2324
%	(1981).


\end{thebibliography}
\end{document}